\documentclass[aps,prl,reprint,superscriptaddress,nofootinbib,floatfix]{revtex4-1}
\usepackage{amsmath,amssymb,bm}
\usepackage{booktabs}
\usepackage{graphicx}
\usepackage{microtype}
\usepackage[T1]{fontenc}
\usepackage{lmodern}
\usepackage{xcolor}
\usepackage[colorlinks=true,hypertexnames=false]{hyperref}

\hypersetup{
  pdftitle={Single-Phonon-Gated Pair Charging of Quantum Batteries},
  pdfauthor={Xiaolong Zhao, Yazhuang Miao, Shuo Liang, Tianhui Qiu, and Libin Fu},
  citecolor=blue,
  linkcolor=red,
  urlcolor=blue
}

\begin{document}

\title{Single-Phonon-Gated Pair Charging of Quantum Batteries}

\author{Xiaolong Zhao}
\affiliation{School of Science, Qingdao University of Technology, Qingdao, Shandong, China}
\author{Yazhuang Miao}
\affiliation{School of Science, Qingdao University of Technology, Qingdao, Shandong, China}
\author{Shuo Liang}
\affiliation{School of Electronic Engineering, Xidian University,  Xi'an 710071, China}
\author{Tianhui Qiu}
\affiliation{School of Science, Qingdao University of Technology, Qingdao, Shandong, China}
\author{Libin Fu}
\email{lbfu@gscaep.ac.cn}
\affiliation{Graduate School, China Academy of Engineering Physics, Beijing 100193, China}

\date{\today}

\begin{abstract}
Charging a quantum battery requires both an energy source and access
to a channel through which that energy can be transferred.
If that access depends on a quantum state, charging becomes conditional
on the state while the external controls remain unchanged.
We show that a single phonon in a common mechanical mode enables a
second-order resonant pair transition in a two-cell battery formed by the
electronic spins of two nitrogen-vacancy centers, whereas the mechanical
vacuum supports no corresponding resonance.
At fixed controls and mean phonon number, the final stored energy and
ergotropy reached with a one-phonon Fock state are unattainable with
any classical mixture of coherent states.
An energy balance shows that the microwave drives supply nearly all of
the stored energy, while the phonon contributes only a small fraction.
Our results open a route to quantum-state-conditioned charging in
quantum batteries under experimentally feasible parameters.
\end{abstract}

\maketitle

\textit{Introduction.}
Quantum-battery charging intertwines controlled energy transfer with
quantum-state preparation
~\cite{Binder2015,Campaioli2024Review,Ferraro2026Perspective}:
the resulting state fixes both the stored energy and the extractable
work
~\cite{Allahverdyan2004,AlickiFannes2013}.
Interactions and collective effects can reshape these dynamics
~\cite{Le2018,JuliaFarre2020}
and thereby modify the stored energy and extractable work
~\cite{Andolina2019,Shi2022,Erdman2024}.
In suitable regimes, they can also enhance charging power
~\cite{Campaioli2017,Ferraro2018,Gyhm2022,Quach2022}.
Charging can occur only when the driven dynamics connect the
discharged state to higher-energy battery states
~\cite{Andolina2018,Farina2019}.
If this connection depends on the state of a coupled quantum system,
the same control sequence can transfer energy to the battery for one
state of that system while suppressing the transfer for another.
The quantum state then determines whether the protocol charges the
battery rather than merely how charging proceeds.

For a quantum system coupled linearly to a quantized bosonic mode
~\cite{Rabl2009,Blais2021}, transition matrix elements depend on the
boson number~\cite{JaynesCummings1963,Brune1996}, allowing different
Fock sectors to exhibit distinct dynamics~\cite{Rinaldi2025}. When
the same mode couples several quantum systems~\cite{TavisCummings1968,Majer2007},
it can mediate correlated multi-emitter transitions~\cite{Bennett2013,Li2020}.
In the few-quantum regime, even a single bosonic quantum can participate
in such a transition~\cite{Garziano2016,Tomonaga2025}. A collective resonance
that is present in one Fock sector but absent in another under the same drive
would make charging conditional on the bosonic state.

We demonstrate single-phonon-gated pair charging in a two-cell quantum
battery formed by the electronic spins of two nitrogen-vacancy centers
coupled to a common mechanical mode. Under microwave dressing, the dipolar
and spin--phonon couplings jointly induce a second-order pair transition
that is resonant in the one-phonon sector, whereas no corresponding pair
resonance exists in the mechanical vacuum under the same controls.
At the same mean phonon number, the final stored energy and ergotropy
reached with the one-phonon Fock state both exceed a common upper bound
that holds for every classical mixture of coherent
states~\cite{Glauber1963,Sudarshan1963}.
An energy balance shows that work performed by the microwave controls
accounts for nearly all of the stored energy, with only a small
contribution from the phonon.
Together, these results establish quantum-state-conditioned charging:
the state of the bosonic mode determines whether an otherwise
unchanged control sequence charges the battery.

\begin{figure}[t]
\centering
\includegraphics[width=0.96\columnwidth]{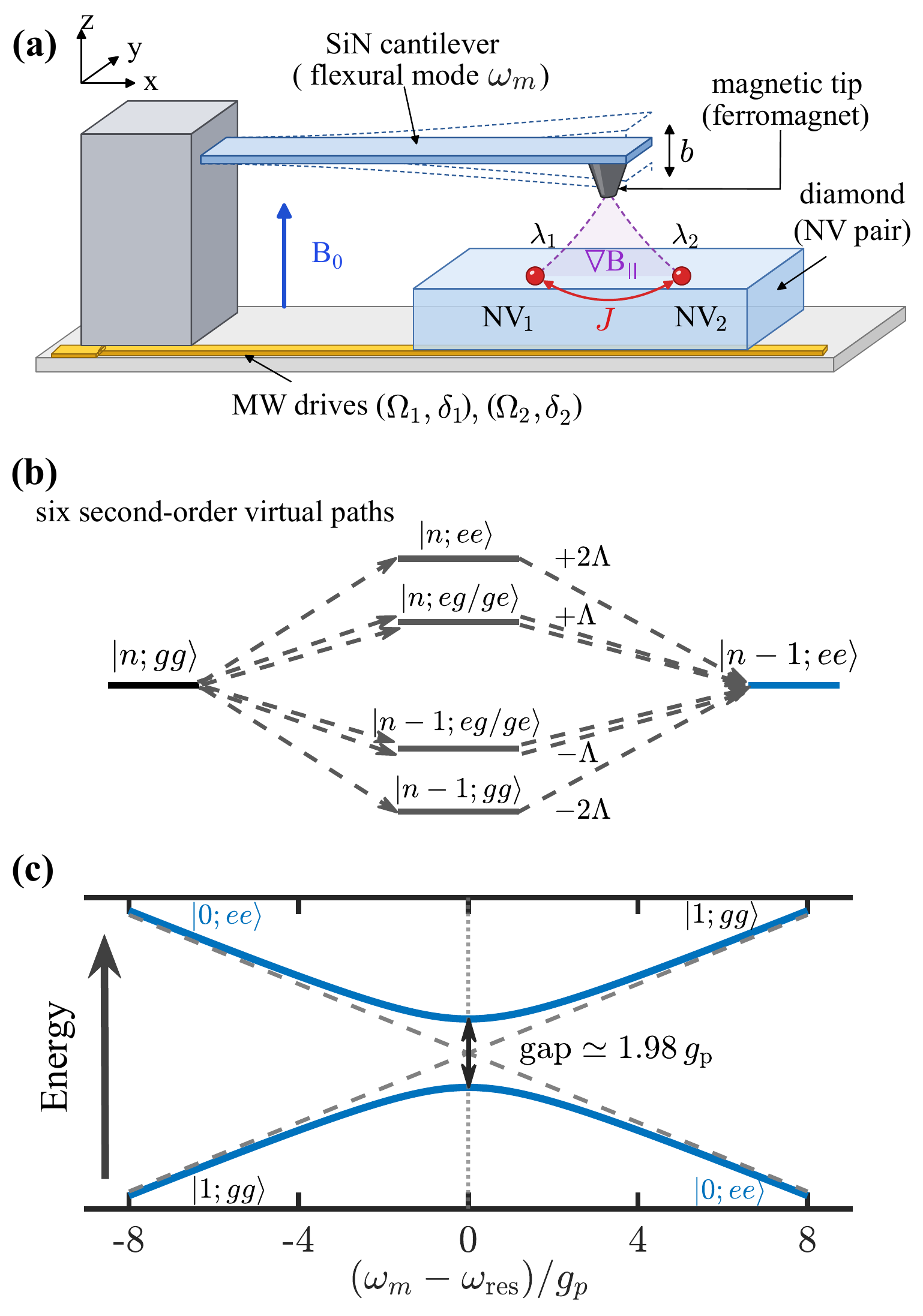}
\caption{(a) Experimental sketch of two microwave-dressed NV-center electronic spins coupled
to a common mechanical mode, with mutual dipolar interaction $J$. (b) Six nonzero second-order
$J\lambda$ virtual paths between $\lvert n;gg\rangle$ and $\lvert n-1;ee\rangle$ near $\omega_m=2\Lambda$.
(c) Pair-dominated eigenenergies of Eq.~\eqref{eq:centralH} (solid blue lines) versus
$(\omega_m-\omega_{\rm res})/g_{\rm p}$, with $\omega_{\rm res}$ the projected one-phonon
avoided-crossing center and $g_{\rm p}$ the reference pair-coupling scale. Gray dashed lines
denote the diabatic pair-state energies centered at $\omega_{\rm res}$; the double arrow denotes
the minimum avoided-crossing gap.}
\label{fig:mechanism}
\end{figure}

\textit{Single-phonon-gated pair charging.}
We consider two NV-center electronic spins (cells) with spectrally resolved transitions and
encode the discharged and charged states of cell $j$ ($j=1,2$) as $\lvert0_j\rangle\equiv\lvert m_s=0\rangle_j$
and $\lvert1_j\rangle\equiv\lvert m_s=-1\rangle_j$
\cite{Song2025SelfDischarge,Doherty2013,PRL136200802,Neumann2010,Yamamoto2013}.
Both spins couple longitudinally to a common mechanical mode of frequency $\omega_m$ through
axial magnetic-field gradients with coupling strengths $\lambda_j$~\cite{Li2020}. The two
electron spins also interact through a magnetic dipole--dipole coupling of strength
$J$~\cite{Doherty2013,PRL136200802,Neumann2010,Yamamoto2013}. Independent microwaves drive
the encoded transitions with detunings $\delta_j$ and Rabi frequencies $\Omega_j$~\cite{SupplementalMaterial}.
The resulting system is illustrated in Fig.~\ref{fig:mechanism}(a). We transform the encoded
spins to their respective microwave rotating frames, apply the rotating-wave approximation (RWA) to the
microwave drives, and impose the secular approximation on the dipolar interaction~\cite{SupplementalMaterial}.
The resulting Hamiltonian reads $H_{\rm enc}/\hbar=\omega_m b^\dagger b +\sum_j(\delta_j\tau_j^z+\Omega_j\tau_j^x)/2
+\sum_j\lambda_j(b+b^\dagger)\tau_j^z +J\tau_1^z\tau_2^z$. Here $b$ annihilates a phonon of
the mechanical mode, $\tau_j^z=\lvert1_j\rangle\langle1_j\rvert -\lvert0_j\rangle\langle0_j\rvert$,
and $\tau_j^x=\lvert0_j\rangle\langle1_j\rvert+\mathrm{H.c.}$ act in the encoded spin subspace.

Diagonalizing the local driven-spin term (excludes $J\tau_1^z\tau_2^z$ in $H_{\rm enc}/\hbar$)
defines the dressed states $\lvert g_j\rangle$ and $\lvert e_j\rangle$ with splitting
$\Lambda_j=(\delta_j^2+\Omega_j^2)^{1/2}$. We define Pauli operations in the dressed basis
$\sigma_j^z\lvert g_j\rangle=-\lvert g_j\rangle$ and $\sigma_j^z\lvert e_j\rangle=\lvert e_j\rangle$
for the $j$th spin. Introducing the mixing angle $\theta_j$ through $\cos\theta_j=\delta_j/\Lambda_j$
and $\sin\theta_j=\Omega_j/\Lambda_j$, the operator $\tau_j^z$ transforms as
$\tau_j^z=\cos\theta_j\,\sigma_j^z-\sin\theta_j\,\sigma_j^x
\equiv A_j$, where $\sigma_j^{x}$ is the Pauli-$x$ operator. Then one obtains the projected
Hamiltonian as
\begin{equation}
\frac{H}{\hbar}
=
\omega_m b^\dagger b
+\sum_{j=1}^{2}\frac{\Lambda_j}{2}\sigma_j^z
+\sum_{j=1}^{2}\lambda_j(b+b^\dagger)A_j
+J A_1A_2 .
\label{eq:centralH}
\end{equation}
We first analyze the symmetric case
$\Lambda_1=\Lambda_2=\Lambda$,
$\theta_1=\theta_2=\theta$, and
$\lambda_1=\lambda_2=\lambda$.
We write $\lvert n;xy\rangle\equiv \lvert n\rangle_m\otimes\lvert x_1y_2\rangle$, where
$\lvert n\rangle_m$ is the $n$-phonon Fock state of the mechanical mode, $n=0,1,\ldots$,
and $x,y\in\{g,e\}$. For $n\ge1$, the energies of $\lvert n;gg\rangle$ and $\lvert n-1;ee\rangle$
differ by $\hbar(2\Lambda-\omega_m)$ and coincide at the pair resonance  $\omega_m=2\Lambda$.
But Eq.~\eqref{eq:centralH} has no direct matrix element between these states. The coupling
between these two dressed states appears indirectly at order $J\lambda$ through virtual states
detuned by $O(\Lambda)$~\cite{SupplementalMaterial}.

To characterize this indirect pair coupling quantitatively, we identify six nonzero
second-order $J\lambda$ paths through the virtual manifolds shown in Fig.~\ref{fig:mechanism}(b).
At the pair resonance $\omega_m=2\Lambda$, all six contributions have equal magnitude.
The four paths through states with one dressed spin excited add constructively, whereas
the remaining two enter with the opposite sign and partially cancel the combined
contribution of the four paths~\cite{SupplementalMaterial}. A second-order Schrieffer--Wolff
elimination of the virtual states~\cite{Bravyi2011} then gives the effective pair Hamiltonian
\begin{equation}
\begin{aligned}
\frac{H_{{\rm p},n}}{\hbar}
&=\frac{\delta_{{\rm p},n}}{2}\eta_z+g_n\eta_x,\\
g_n(\Lambda,\theta)
&=\sqrt n\,g_{\rm pair}(\Lambda,\theta).
\end{aligned}
\label{eq:pairH}
\end{equation}
Here $\eta_z=\lvert n-1;ee\rangle\langle n-1;ee\rvert -\lvert n;gg\rangle\langle n;gg\rvert$
and $\eta_x=\lvert n-1;ee\rangle\langle n;gg\rvert+\mathrm{H.c.}$ act in the pair subspace
$\{\lvert n;gg\rangle,\lvert n-1;ee\rangle\}$. The pair-coupling coefficient is
$g_{\rm pair}(\Lambda,\theta) = 2\frac{J\lambda}{\Lambda}\cos\theta\sin^2\theta$,
the factor $\sqrt n$ follows directly from $\langle n-1|b|n\rangle=\sqrt n$, details in
the Supplemental Material~\cite{SupplementalMaterial} S2. Near the unperturbed pair
resonance, the detuning entering Eq.~\eqref{eq:pairH} is, through second order,
$\delta_{{\rm p},n} \simeq 2\Lambda-\omega_m +\frac{J^2}{\Lambda}
\left(4\cos^2\theta\sin^2\theta+\sin^4\theta\right) -\frac{8n\lambda^2}{3\Lambda}\sin^2\theta$
~\cite{SupplementalMaterial}. For $0\leq\theta\leq\pi/2$, the angular factor
$\cos\theta\sin^2\theta$ is maximal at $\cos^2\theta=1/3$, giving $\theta_0=\arccos(1/\sqrt3)$.
We denote by $\Lambda_0$ the reference dressed splitting used in the following,
chosen in the perturbative regime $|J|,|\lambda|\ll\Lambda_0$, and define
$g_{\rm p}\equiv|g_{\rm pair}(\Lambda_0,\theta_0)|$. For the numerical examples,
$(J,\lambda,\Lambda_0)/2\pi=(53.0,6.00,700)~\mathrm{kHz}$, here the choice of $J$
is consistent, after projection onto the encoded subspace, with dipolar interactions
measured in strongly coupled NV pairs~\cite{Lee2023}, and $\lambda/2\pi=6.00~\mathrm{kHz}$
is below the $\sim100~\mathrm{kHz}$ magnetic-gradient coupling estimated for a
nanomechanical NV device~\cite{Li2020}. With $\Lambda_0/2\pi=700~\mathrm{kHz}$,
the ratios $J/\Lambda_0\simeq0.076$ and $\lambda/\Lambda_0\simeq8.6\times10^{-3}$
remain perturbative; the required microwave-dressing scale is also below MHz-scale
Rabi frequencies demonstrated for single NV centers~\cite{Golter2013}.

For the one-phonon sector ($n=1$), direct diagonalization of Eq.~\eqref{eq:centralH}
at the reference point $(\Lambda_0,\theta_0)$ yields an isolated avoided crossing
between the pair-dominated branches associated with $\lvert1;gg\rangle$ and $\lvert0;ee\rangle$.
We denote the center of this projected one-phonon avoided crossing by $\omega_{\rm res}$
and show the spectrum in Fig.~\ref{fig:mechanism}(c). The two branches exchange their
pair-state character across the crossing while retaining nearly unit weight in the pair
subspace, and the avoided-crossing half-gap agrees with $g_{\rm p}$ at the percent
level~\cite{SupplementalMaterial}. At the same reference point, $\delta_{{\rm p},1}=0$
in Eq.~\eqref{eq:pairH} defines the second-order one-phonon resonance
$\omega_{\rm res}^{(2)}\simeq 2\Lambda+\frac{\sin^2\theta}{3\Lambda}\left[3J^2(1+3\cos^2\theta)-8\lambda^2\right]$.
The small difference between $\omega_{\rm res}^{(2)}$ and $\omega_{\rm res}$ arises
from higher-order corrections. For the one-phonon charging protocol with initial state
$\lvert1;gg\rangle$, setting $\omega_m=\omega_{\rm res}^{(2)}$ in Eq.~\eqref{eq:pairH}
gives the double-excitation probability $P_{ee}^{\rm eff}(t)=\sin^2(g_{\rm p}t)$, with
$t_\pi=\pi/(2g_{\rm p})$.

\begin{figure}[t]
\centering
\includegraphics[width=0.96\columnwidth]{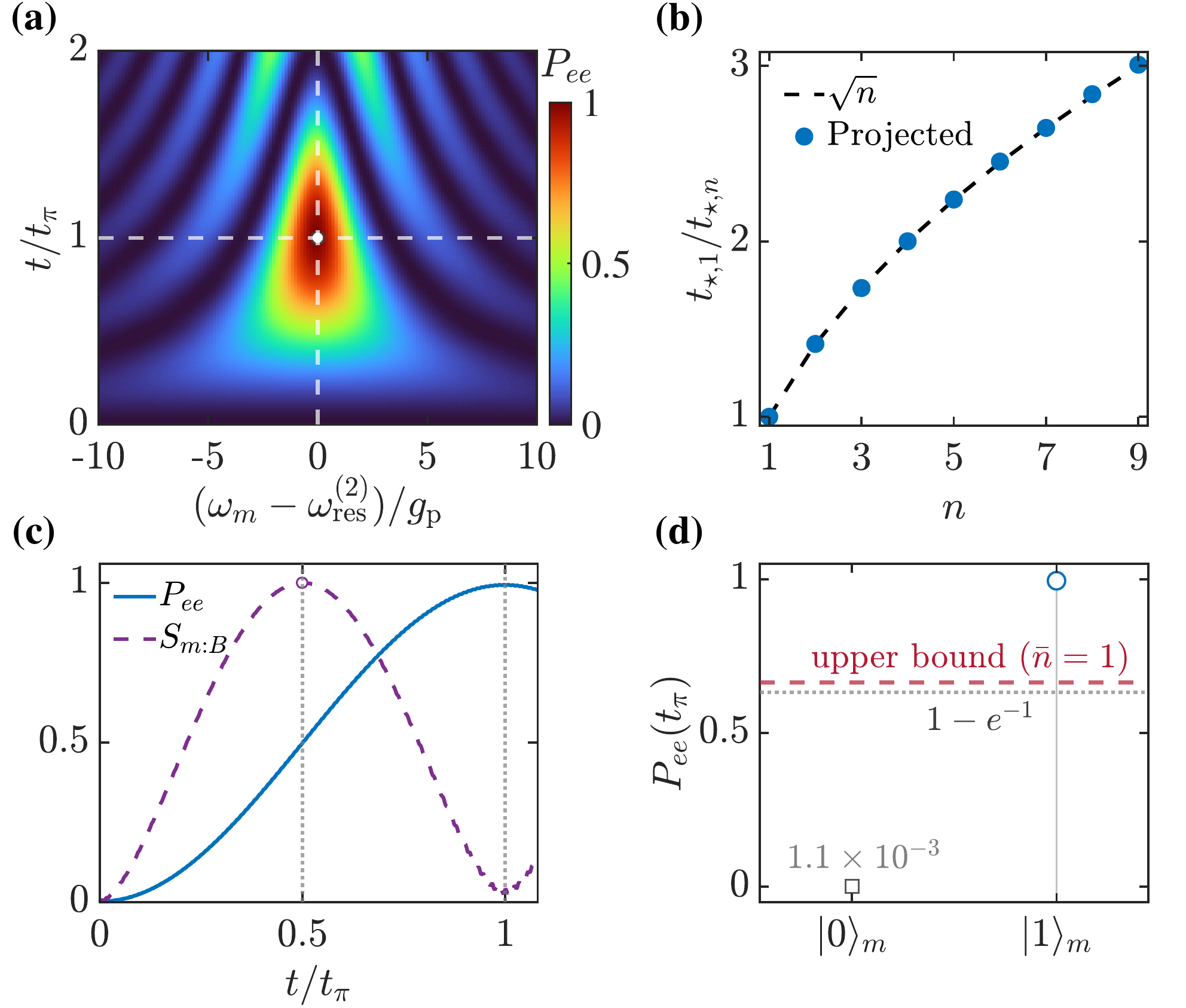}
\caption{ (a) Dressed double-excitation population $P_{ee}$ for the one-phonon input as a
function of $(\omega_m-\omega_{\rm res}^{(2)})/g_{\rm p}$ and $t/t_\pi$. Dashed lines denote
$\omega_m=\omega_{\rm res}^{(2)}$ and $t=t_\pi$. (b) Values from the projected model for
$t_{\star,1}/t_{\star,n}$ at the projected avoided-crossing center of each Fock sector
(markers), together with the $\sqrt n$ dependence (dashed line). (c) Dressed double-excitation
population $P_{ee}$ and mode--battery entanglement entropy $S_{m:B}$ for the one-phonon input
at $\omega_m=\omega_{\rm res}^{(2)}$. Dotted vertical lines denote $t=t_\pi/2$ and $t=t_\pi$.
(d) $P_{ee}(t_\pi)$ for vacuum and one-phonon Fock inputs under the same controls at
$\omega_m=\omega_{\rm res}^{(2)}$. The red dashed line denotes the $\bar n=1$ classical upper
bound for states with nonnegative-$P_{\rm GS}$; the gray dotted line denotes the ideal vacuum-off
value $1-e^{-1}$.}
\label{fig:gating}
\end{figure}

For the dynamics governed by Eq.~\eqref{eq:centralH}, let $\rho(t)$ denote the joint
spin--mechanical density operator. We define the dressed double-excitation population
as $P_{ee}(t)\equiv \sum_{\ell=0}^{\infty} \langle \ell;ee|\rho(t)|\ell;ee\rangle$,
where $\ell$ labels the mechanical Fock basis. Direct evolution from $\lvert1;gg\rangle$
at $\omega_m=\omega_{\rm res}^{(2)}$ brings $P_{ee}$ close to unity near $t_\pi$ and
then produces coherent back-transfer. The behavior of $P_{ee}$ versus mechanical
frequency and time is shown in Fig.~\ref{fig:gating}(a). At the reference dressed controls
$(\Lambda_0,\theta_0)$, Eq.~\eqref{eq:pairH} gives $|g_n|=\sqrt n\,g_{\rm p}$ for $n\ge1$,
with the corresponding resonant transfer time $t_\pi/\sqrt n$. In the projected model, $t_{\star,n}$
is the time defined at the first maximum of $P_{ee}(t)$ at the projected avoided-crossing
center of each Fock sector, and the weak $n$-dependent resonance shifts do not affect
the comparison. The ratios $t_{\star,1}/t_{\star,n}$ follow the predicted $\sqrt n$
dependence, as shown in Fig.~\ref{fig:gating}(b). Thus higher phonon numbers lead to shorter
resonant transfer times.

Naturally, the coherent pair exchange also generates transient entanglement between the
mechanical mode and the two-spin battery. For a resonant pair transition in Fock sector
$n$, Eq.~\eqref{eq:pairH} gives $\lvert\Psi_n(t)\rangle = \cos(g_nt)\lvert n;gg\rangle
-i\sin(g_nt)\lvert n-1;ee\rangle$. With $\rho_B(t)=\operatorname{Tr}_m\rho(t)$, we quantify
the mechanical mode--battery entanglement by the entropy $S_{m:B}(t)=-\operatorname{Tr}[\rho_B(t)\log_2\rho_B(t)]$.
For the state $\lvert\Psi_n(t)\rangle$ above, $S_{m:B}(t)=h_2[\sin^2(g_nt)]$, where
$h_2(p)=-p\log_2p-(1-p)\log_2(1-p)$ is the binary entropy ~\cite{Shannon1948}. In the
projected one-phonon dynamics shown by Fig.~\ref{fig:gating}(c), $S_{m:B}$ rises during
the pair transfer and becomes small again as $P_{ee}$ reaches its first maximum near $t_\pi$,
consistent with the effective two-state result $S_{m:B}(t)=h_2[\sin^2(g_1t)]$. The underlying
oscillatory exchange is characteristic of resonant Jaynes--Cummings dynamics~\cite{JaynesCummings1963,PhoenixKnight1991},
while a similar behavior of rise and reduction of entanglement during population transfer
has been identified in quantum-battery charging~\cite{Shi2022}.

The mechanical vacuum provides the gate-off state, as every $\lvert\ell;ee\rangle$
lies $\hbar(\ell\omega_m+2\Lambda)$ above $\lvert0;gg\rangle$, precluding a resonant
pair transition. Equation~\eqref{eq:pairH} couples $\lvert n;gg\rangle$ to the
lower-phonon partner $\lvert n-1;ee\rangle$, so $n=1$ is the lowest Fock sector that
supports the resonant pair transition.
Off-resonant couplings retained in Eq.~\eqref{eq:centralH}
nevertheless produce a small but nonzero vacuum response
$P_{ee,{\rm vac}}(t_\pi)$, shown in Fig.~\ref{fig:gating}(d).

The gate-off effect resulting from vacuum sets an upper bound on the pair-transfer
probability attainable from classical mechanical states with the same mean phonon
number, $\bar n=1$. We define such states by a nonnegative Glauber--Sudarshan
representation $\rho_{\rm cl}=\int d^2\alpha\,P_{\rm GS}(\alpha)
\lvert\alpha\rangle_m{}_m\!\langle\alpha\rvert$, with
$P_{\rm GS}(\alpha)\ge0$~\cite{Glauber1963,Sudarshan1963}.
The mean-occupation constraint is $\int d^2\alpha\,P_{\rm GS}(\alpha)|\alpha|^2=1$,
while the vacuum weight $p_0\equiv{}_m\!\langle0|\rho_{\rm cl}|0\rangle_m
=\int d^2\alpha\,P_{\rm GS}(\alpha)e^{-|\alpha|^2}$ obeys
$p_0\ge e^{-1}$ by Jensen's inequality, whereas $p_0=0$ for the one-phonon Fock
state $\lvert1\rangle_m$. With the Hamiltonian, initial spin state, and readout
time fixed, only the mechanical input is varied. Let $P_{ee,{\rm vac}}(t)$ and
$P_{ee,{\rm cl}}(t)$ denote the responses for the vacuum and a classical
state as inputs, respectively. At the common readout time $t_\pi$, we write
$P_{ee,{\rm vac}}\equiv P_{ee,{\rm vac}}(t_\pi)$ and $P_{ee,{\rm cl}}\equiv P_{ee,{\rm cl}}(t_\pi)$.
The vacuum weight constrains the classical response through the fidelity between
the vacuum and classical inputs. The root fidelity between the vacuum state and
$\rho_{\rm cl}$ is $\sqrt{p_0}$ which is preserved by the unitary evolution, and
the final binary measurement associated with $P_{ee}$ cannot decrease it
~\cite{SupplementalMaterial}. Hence
$\sqrt{P_{ee,{\rm cl}}P_{ee,{\rm vac}}}
+\sqrt{(1-P_{ee,{\rm cl}})(1-P_{ee,{\rm vac}})}
\ge\sqrt{p_0}\ge e^{-1/2}$.
For $P_{ee,{\rm vac}}<e^{-1}$, maximizing the allowed classical
response gives~\cite{SupplementalMaterial}
\begin{equation}
P_{ee,{\rm cl}}\le
\left[
\sqrt{e^{-1}P_{ee,{\rm vac}}}
+\sqrt{(1-e^{-1})(1-P_{ee,{\rm vac}})}
\right]^2.
\label{eq:classicalBound}
\end{equation}
In the ideal vacuum gate-off limit, the bound reduces to $1-e^{-1}$. The finite vacuum
response raises the bound slightly, but the one-phonon response $P_{ee}(t_\pi)$ remains
above it, as shown in Fig.~\ref{fig:gating}(d). Thus, under the same operation and at
fixed mean phonon number $\bar n=1$, the one-phonon Fock state yields a pair-transfer
response unattainable by any classical mechanical state.

\textit{From pair transfer to stored work.}
The pair transfer mentioned above does not yet leave the battery in a drive-free charged
state. At $t_\pi$, the population resides predominantly in $\lvert0;ee\rangle$, but
$\lvert ee\rangle$ remains microwave dressed and resonant pair coupling can drive
back-transfer to $\lvert1;gg\rangle$, as seen for $t>t_\pi$ in Fig.~\ref{fig:gating}(a),
a familiar limitation of coherent quantum-battery charging~\cite{Santos2019Stable}.
The desired endpoint is instead the bare charged state $\lvert11\rangle$, reached by
suppressing reverse exchange and removing the microwave dressing.

From the reduced battery state $\rho_B(t)$, we define $P_{11}(t)=\langle11|\rho_B(t)|11\rangle$
to track the population of $\lvert11\rangle$ during the protocol. We quantify the stored
energy and ergotropy using the bare-spin Hamiltonian $H_B=\hbar\omega_{B,1}N_1+\hbar\omega_{B,2}N_2
+4\hbar JN_1N_2$, where $\omega_{B,j}$ are the drive-free bare-spin transition frequencies
and $N_j=\lvert1_j\rangle\langle1_j\rvert$. We take
$(\omega_{B,1},\omega_{B,2})/2\pi=(2.59975,2.57975)~\mathrm{GHz}$ with the $20~\mathrm{MHz}$
separation being much larger than the reference dressing scale $\Lambda_0/2\pi=700~\mathrm{kHz}$
and is consistent with frequency-selective addressing of coupled NV centers
~\cite{Neumann2010,Yamamoto2013}. The battery energy and capacity are
$E_B(t)=\operatorname{Tr}[\rho_B(t)H_B]$ and $E_{\rm cap}=\langle11|H_B|11\rangle-\langle00|H_B|00\rangle$,
while the ergotropy is $\mathcal W(t)=E_B(t)-\operatorname{Tr}[\pi_B(t)H_B]$, where $\pi_B(t)$
is the passive state associated with $\rho_B(t)$ and $H_B$~\cite{Allahverdyan2004,Francica2020,Niu2024}.
Throughout the driven sequence, $E_B(t)$ and $\mathcal W(t)$ are evaluated with this fixed
$H_B$; after the microwave drives are switched off, they represent the stored battery energy
and extractable work.

\begin{figure}[t]
\centering
\includegraphics[width=0.94\columnwidth]{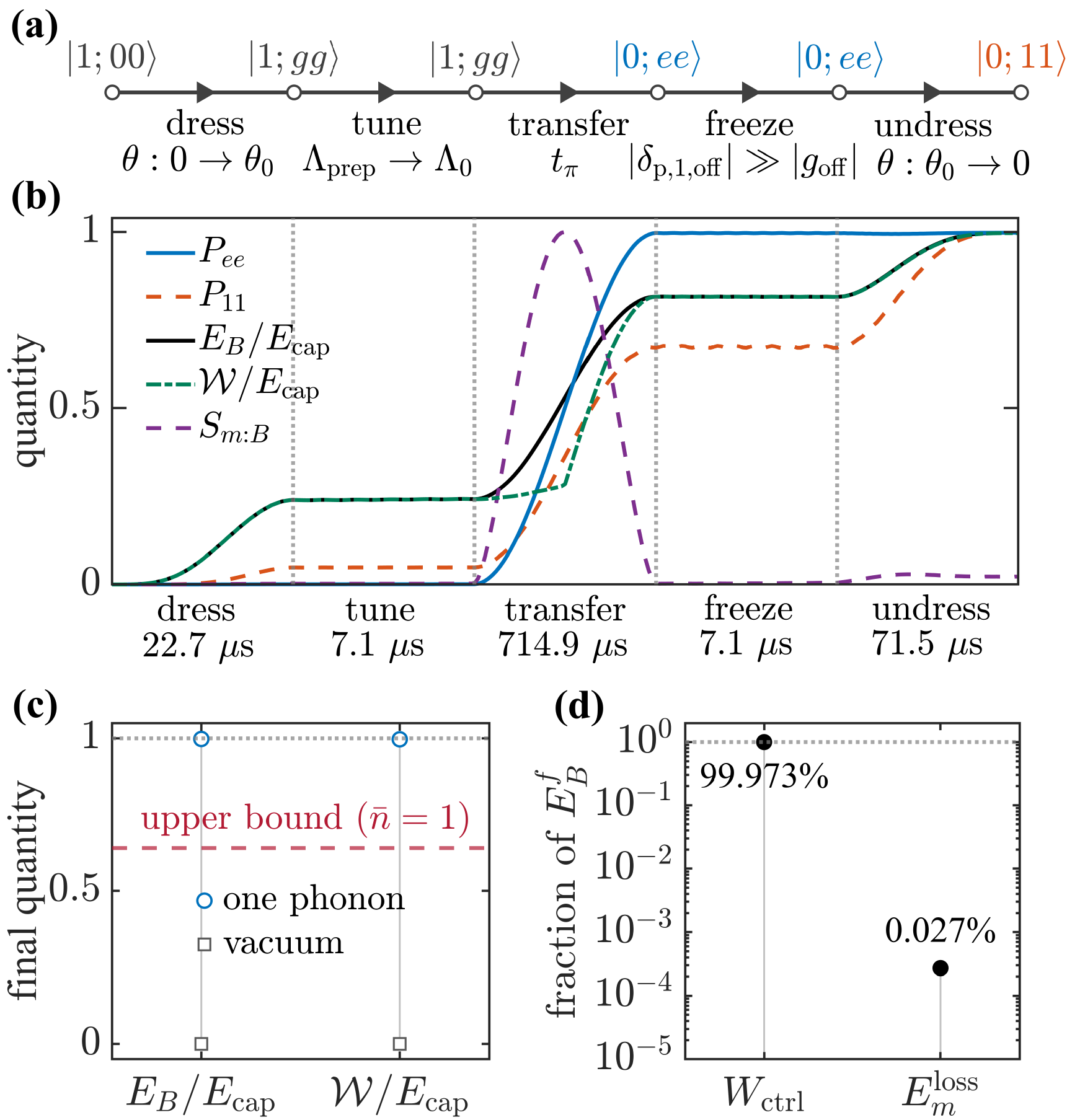}
\caption{(a) Five-stage charging sequence: dress, tune, transfer, freeze, and undress. The
mechanical frequency is fixed at $\omega_{\rm res}^{(2)}$, and the transfer stage lasts
$t_\pi=\pi/(2g_{\rm p})$. $\Lambda_{\rm prep}$ denotes the preparation-stage dressed splitting;
$\delta_{{\rm p},1,{\rm off}}$ and $g_{\rm off}$ denote the pair detuning and pair-coupling
coefficient at the off-resonant freeze point. (b) One-phonon evolution of the dressed
double-excitation population $P_{ee}$, bare charged-state population $P_{11}$, battery energy
$E_B/E_{\rm cap}$, ergotropy $\mathcal W/E_{\rm cap}$, and mode--battery entanglement entropy
$S_{m:B}$. The five stages are shown with equal graphical widths (time-scaled abscissa), with
their physical durations indicated below. (c) Final $E_B/E_{\rm cap}$ and $\mathcal W/E_{\rm cap}$
for one-phonon and vacuum inputs under the same controls. The red dashed line denotes the full-cycle
classical upper bound for mechanical inputs with $\bar n=1$. (d) Control work $W_{\rm ctrl}$
and mechanical-energy loss $E_m^{\rm loss}$, normalized by the final one-phonon battery energy
$E_B^f$.}
\label{fig:battery}
\end{figure}

During the control sequence, $P_{ee}$ is evaluated in the instantaneous dressed basis. For bare
encoded labels $a,b\in\{0,1\}$, we write $\lvert n;ab\rangle\equiv \lvert n\rangle_m\otimes\lvert a_1b_2\rangle$.
Initially, $\theta=0$, $\Omega_j=0$ and $\tau_j^z=\sigma_j^z$ for $j=1,2$, and the initial state
$\lvert1;00\rangle$ coincides with $\lvert1;gg\rangle$. The mechanical frequency is fixed at
$\omega_m=\omega_{\rm res}^{(2)}$ throughout the sequence, and all ramps use the cubic smoothstep
$\mathcal R(u)=3u^2-2u^3$, $0\le u\le1$, for which the ramp velocity vanishes at both endpoints
~\cite{SupplementalMaterial}. Dress keeps $\Lambda=\Lambda_{\rm prep}$ and ramps $\theta:0\to\theta_0$
adiabatically. The preparation splitting $\Lambda_{\rm prep}$ is chosen so that
$|\delta_{{\rm p},1}|\gg|g_{\rm pair}(\Lambda_{\rm prep},\theta)|$ throughout the ramp, suppressing
premature pair exchange; the charging protocol in Fig.~\ref{fig:battery} uses $\Lambda_{\rm prep}=1.10\Lambda_0$,
with $|\dot\theta|\ll\Lambda_{\rm prep}$ for local adiabatic following~\cite{SupplementalMaterial}.
Then in the tune stage, $\theta=\theta_0$ and $\Lambda$ varies as: $\Lambda_{\rm prep}\to\Lambda_0$,
bringing the one-phonon pair transition to its second-order resonance. Since varying $\Lambda$ at
fixed $\theta$ leaves the dressed basis unchanged~\cite{Doherty2013,PRL136200802,Golter2013}, the
tune stage can be short on the pair-exchange timescale. The control parameters are held at
$(\Lambda_0,\theta_0)$ for a duration $t_\pi$; during this transfer stage $P_{ee}$ rises while
the entropy $S_{m:B}$ captures the transient mechanical mode--battery entanglement shown in
Fig.~\ref{fig:battery}(b). Then the freeze stage keeps $\theta=\theta_0$ and shifts $\Lambda$ to
the off-resonant value $\Lambda_{\rm off}$, avoiding pair exchange back to $\lvert1;gg\rangle$.
At this freeze onset, we define $g_{\rm off}\equiv g_{\rm pair}(\Lambda_{\rm off},\theta_0)$ and
denote by $\delta_{{\rm p},1,{\rm off}}$ the detuning $\delta_{{\rm p},1}$ of Eq.~\eqref{eq:pairH},
which gives the reverse-transfer envelope $4g_{\rm off}^2/[\delta_{{\rm p},1,{\rm off}}^2+4g_{\rm off}^2]$.
We choose $\Lambda_{\rm off}=\Lambda_0+20g_{\rm p}$ so that $|\delta_{{\rm p},1,{\rm off}}|\gg|g_{\rm off}|$,
suppressing back-transfer after the population exchange~\cite{SupplementalMaterial}. Undress then
keeps $\Lambda=\Lambda_{\rm off}$ and adiabatically ramps $\theta:\theta_0\rightarrow0$ while the
pair transition remains off resonance. As $\theta\rightarrow0$, $\Omega_j\rightarrow0$ and
$\tau_j^z\rightarrow\sigma_j^z$, and the dressed state $\lvert0;ee\rangle$ evolves into the bare
charged state $\lvert0;11\rangle$. Correspondingly, $P_{11}$, $E_B/E_{\rm cap}$, and $\mathcal W/E_{\rm cap}$
approach unity while $S_{m:B}$ becomes small, as shown in Fig.~\ref{fig:battery}(b). We use
$\tau_{\rm dress}=100/\Lambda_0$, $\tau_{\rm tune}=\tau_{\rm freeze}=0.01t_\pi$, and
$\tau_{\rm undress}=0.10t_\pi$; dress and undress are slow on the local dressed-state scale, while
tune and freeze are short compared to the pair-exchange scale~\cite{SupplementalMaterial}. The
control sequence is summarized in Fig.~\ref{fig:battery}(a). After undress, the remaining longitudinal
spin--mechanical interaction commutes with $H_B$ and therefore cannot change $E_B$. The same five-stage
control sequence is used for all mechanical-input comparisons.

The fidelity between vacuum and classical state argument leading to Ineq.~\eqref{eq:classicalBound} also
constrains the final battery energy. Here $\lvert00\rangle$ is the lowest-energy eigenstate of $H_B$
with energy $0$, while $\lvert11\rangle$ is the highest-energy eigenstate with energy $E_{\rm cap}$.
Writing $\mathbb I_B$ for the identity on the battery Hilbert space, $0\le H_B/E_{\rm cap}\le\mathbb I_B$,
so Ineq.~\eqref{eq:classicalBound} about population also applies to $E_B^f/E_{\rm cap}$, where $E_{B}^f$
denote the final battery energy for different kinds of input along with the nonnegative-$P_{\rm GS}$ upper
bound is shown in Fig.~\ref{fig:battery}(c)~\cite{SupplementalMaterial}. The one-phonon value of
$E_B^f/E_{\rm cap}$ exceeds this bound, whereas the vacuum input produces negligible final battery
energy under the same controls, as shown in Fig.~\ref{fig:battery}(c). Since $\mathcal W^f\le E_B^f$,
the same bound also upper-bounds $\mathcal W^f/E_{\rm cap}$, which is exceeded by the one-phonon result.
Thus, under the same five-stage controls and at fixed mean phonon number $\bar n=1$, the final stored
energy and ergotropy produced by the one-phonon Fock input exceed the upper bound for any classical
mechanical input.

The single phonon gates the resonant pair-transfer channel; whether it also supplies the energy stored
in the battery is another question. Physically, for the closed driven system, the change in total
laboratory-frame energy is the work performed by the time-dependent microwave controls,
$W_{\rm ctrl}=E_{\rm lab}(t_f)-E_{\rm lab}(0)$, where $E_{\rm lab}(t)$ is the total laboratory-frame energy
and $t_f$ is the end of the control sequence. We write $E_m(t)=\hbar\omega_m\langle b^\dagger b\rangle_t$
for the mechanical energy and $E_{\rm int}(t)$ for the longitudinal spin--mechanical interaction energy
at time $t$. The microwave drives vanish at $t=0$ and after $t_f$, and the initial state $\lvert1;00\rangle$
gives $E_B(0)=E_{\rm int}(0)=0$. With $E_m^{\rm loss}=E_m(0)-E_m(t_f)$, energy conservation gives
\begin{equation}
E_B^f + E_{\rm int}^f=W_{\rm ctrl}+E_m^{\rm loss},
\label{eq:energyLedger}
\end{equation}
where $E_B^f=E_B(t_f)$ and $E_{\rm int}^f=E_{\rm int}(t_f)$. $W_{\rm ctrl}$ accounts for nearly all
of the final one-phonon battery energy, while the mechanical-energy loss contributes only a small fraction,
shown in Fig.~\ref{fig:battery}(d). Applying Eq.~\eqref{eq:energyLedger} to the vacuum input, with the same
five-stage control, gives a much smaller $E_B^f$ than for the one-phonon input~\cite{SupplementalMaterial}.
The corresponding battery-energy contrast is dominated by the difference in microwave control work. The phonon
therefore gates the pair-charging channel rather than serving as the dominant energy source.
For the five-stage protocol, the charging efficiency, defined as the ratio of final ergotropy to the sum of
initial phonon energy and accumulated positive microwave work, is $85.15\%$~\cite{SupplementalMaterial}.

\begin{figure}[t]
\centering
\includegraphics[width=0.97\columnwidth]{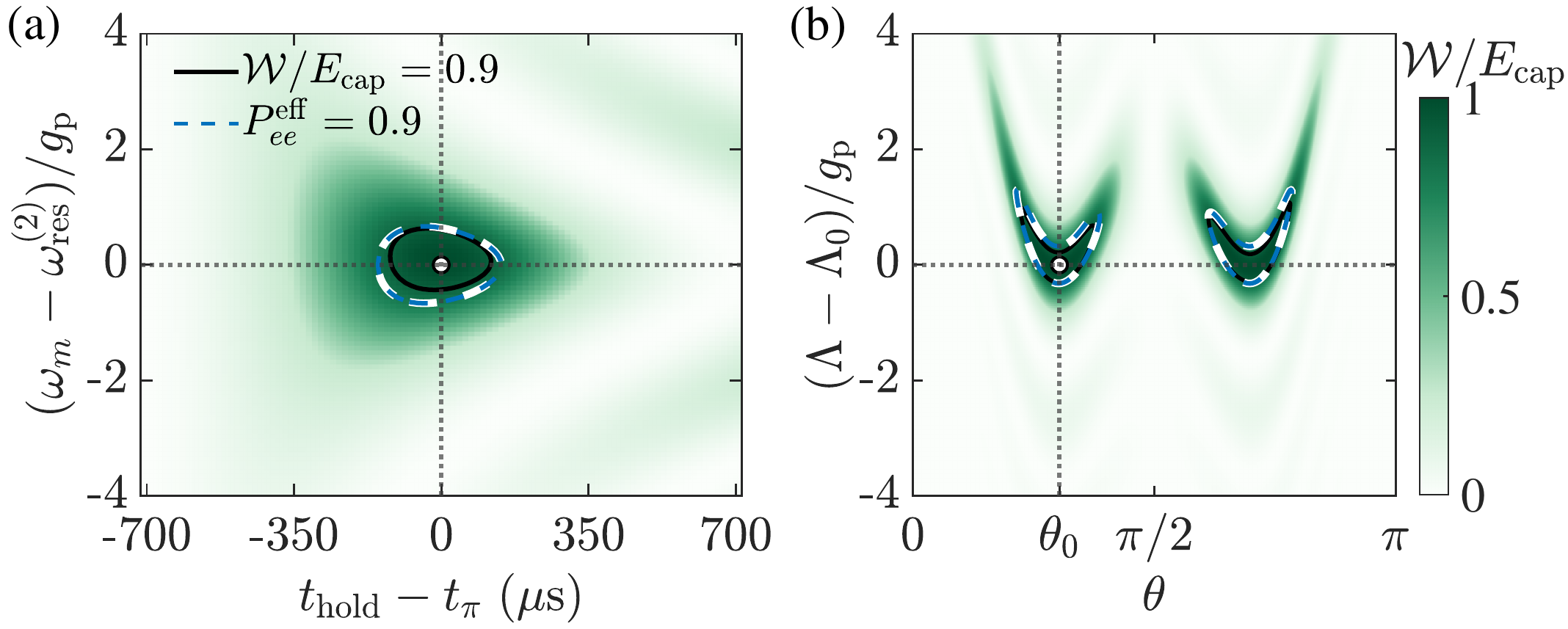}
\caption{
Final one-phonon ergotropy $\mathcal W/E_{\rm cap}$ under variations
of the reference protocol.
(a) Map versus $t_{\rm hold}-t_\pi$ and
$(\omega_m-\omega_{\rm res}^{(2)})/g_{\rm p}$, with the remaining
controls fixed at their reference values.
(b) Map versus the transfer-stage mixing angle $\theta$ and dressed
splitting $(\Lambda-\Lambda_0)/g_{\rm p}$, with the corresponding
dressed-control trajectory applied through all five stages and the
reference stage durations retained.
Black solid contours denote $\mathcal W/E_{\rm cap}=0.9$ from the
projected dynamics; blue dashed contours denote
$P_{ee}^{\rm eff}=0.9$ from the effective pair model.}
\label{fig:operating}
\end{figure}

\textit{Detuning and control dependence of resonant pair charging.}
The pair resonance that enables one-phonon charging also constrains the variation ranges of mechanical
frequency and transfer time over which the reference protocol retains high final ergotropy. For the one-phonon
transfer stage in Fig.~\ref{fig:battery}(b), $P_{ee}$ can be approximated by
$P_{ee}^{\rm eff}(t;\delta_{{\rm p},1},g_{\rm pair}) = 4g_{\rm pair}^2(\delta_{{\rm p},1}^2+4g_{\rm pair}^2)^{-1}
\sin^2[t\sqrt{\delta_{{\rm p},1}^2+4g_{\rm pair}^2}/2]$ obtained from Eq.~\eqref{eq:pairH}.
At $(\Lambda_0,\theta_0)$, $\delta_{{\rm p},1}=-(\omega_m-\omega_{\rm res}^{(2)})$, while on resonance
$P_{ee}^{\rm eff}(t)=\sin^2(g_{\rm p}t) = \cos^2[g_{\rm p}(t-t_\pi)]$. Around $(\omega_{\rm res}^{(2)},t_\pi)$,
these expressions give $P_{ee}^{\rm eff}\simeq 1-(\omega_m-\omega_{\rm res}^{(2)})^2/(4g_{\rm p}^2)
-g_{\rm p}^2(t-t_\pi)^2$. Thus frequency detuning and timing deviations suppress pair transfer quadratically
about $(\omega_m,t) = (\omega_{\rm res}^{(2)},t_\pi)$. As shown in Fig.~\ref{fig:operating}(a), the contour
$P_{ee}^{\rm eff}\ge0.9$ sets $|\omega_m-\omega_{\rm res}^{(2)}|\simeq0.65g_{\rm p}$ at $t=t_\pi$ and
$|t-t_\pi|\simeq0.20t_\pi$ at $\omega_m=\omega_{\rm res}^{(2)}$. The $\mathcal W/E_{\rm cap}=0.9$ contour from the projected dynamics is
in close agreement with the $P_{ee}^{\rm eff}=0.9$ contour in
Fig.~\ref{fig:operating}(a), showing that high final ergotropy
persists over appreciable timing deviations but requires the
mechanical frequency to remain within the pair-coupling scale of
$\omega_{\rm res}^{(2)}$.

We next vary the transfer-stage values of $\Lambda$ and $\theta$ while keeping
$\omega_m=\omega_{\rm res}^{(2)}$ and the stage durations $\tau_{\rm dress}$,
$\tau_{\rm tune}$, $t_\pi$, $\tau_{\rm freeze}$, and $\tau_{\rm undress}$ unchanged.
The corresponding five-stage trajectories $\Lambda(t)$ and $\theta(t)$ are defined
in the Supplemental Material~\cite{SupplementalMaterial}. At $\Lambda=\Lambda_0$,
expanding about $\theta_0$ gives $|g_{\rm pair}(\Lambda_0,\theta)|/g_{\rm p}
=1-3(\theta-\theta_0)^2+O[(\theta-\theta_0)^3]$, so the coupling magnitude changes
only quadratically with $\theta-\theta_0$. At $\theta=\theta_0$, varying $\Lambda$
shifts the pair detuning through the leading term $2\Lambda-\omega_m$, while the
change in $g_{\rm pair}\propto1/\Lambda$ is weak over the range of $\Lambda$
considered in this work. For the effective pair model, $P_{ee}^{\rm eff}\ge0.9$
extends to approximately $|\theta-\theta_0|\lesssim0.15$ at $\Lambda=\Lambda_0$ and
$|\Lambda-\Lambda_0|\lesssim0.32g_{\rm p}$ at $\theta=\theta_0$. The $\mathcal
W/E_{\rm cap}=0.9$ contour versus $(\Lambda-\Lambda_0)/g_p$ and $\theta$ agrees
closely with the $P_{ee}^{\rm eff}=0.9$ contour in Fig.~\ref{fig:operating}(b),
showing that the protocol is substantially more tolerant to variations of $\theta$
than to variations of $\Lambda$ around $(\Lambda_0,\theta_0)$. Moreover,
$g_{\rm pair}(\Lambda,\pi-\theta)=-g_{\rm pair}(\Lambda,\theta)$,
whereas $\delta_{{\rm p},1}(\Lambda,\pi-\theta) =\delta_{{\rm p},1}(\Lambda,\theta)$,
so $P_{ee}^{\rm eff}$ is symmetric about $\theta=\pi/2$. This symmetry gives the two
high-transfer regions near $\theta_0$ and $\pi-\theta_0$ in Fig.~\ref{fig:operating}(b).

\textit{Conclusions.}
We have shown that a single phonon can gate a collective pair-charging channel without
acting as the principal source of the energy stored in the battery. The number-resolved
pair resonance and the nonnegative-$P_{\rm GS}$ upper bound establish that this conditional
access is determined by the quantum state of the mechanical mode rather than by its mean
excitation alone. The selected pair excitation can subsequently be frozen into a drive-free
bare-spin battery. The quantum--classical separation above then persists in the stored
energy, which is almost entirely extractable as ergotropy. The laboratory-frame balance
further separates pathway selection from energy supply by identifying the microwave controls
as the dominant source of the stored energy with experimentally feasible parameters.

\par\vspace{1\baselineskip}
\emph{Acknowledgments.}
This work was supported by the Joint Fund of Natural Science Foundation
of Shandong Province (Grant No. ZR2024LLZ004), the National Natural
Science Foundation of China (Grant Nos. 12575010 and 12005110), the
Natural Science Foundation of Shandong Province
(Grant Nos. ZR2020QA078 and ZR2023MD064), and the Youth Innovation
Team Program of Shandong Province (Grant No. 2023KJ118).

\par\vspace{1\baselineskip}
\emph{Data Availability.}
The data that support the findings of this article are openly
available~\cite{dataset_Miao}, embargo periods may apply.

\clearpage
\onecolumngrid

\setcounter{equation}{0}
\renewcommand{\theequation}{S\arabic{equation}}
\renewcommand{\theHequation}{supp.eq.\arabic{equation}}
\setcounter{table}{0}
\renewcommand{\thetable}{S\arabic{table}}
\renewcommand{\theHtable}{supp.tab.\arabic{table}}
\setcounter{section}{0}
\setcounter{secnumdepth}{2}
\renewcommand{\thesection}{S\arabic{section}}
\renewcommand{\theHsection}{supp.sec.\arabic{section}}
\setcounter{subsection}{0}
\renewcommand{\thesubsection}{\thesection.\Alph{subsection}}
\renewcommand{\theHsubsection}{\theHsection.\Alph{subsection}}
\makeatletter
\renewcommand{\p@subsection}{}
\makeatother

\begin{center}
{\large\bfseries Supplemental Material for\\[3pt]
``Single-Phonon-Gated Pair Charging of Quantum Batteries''}\\[9pt]
Xiaolong Zhao\textsuperscript{1}, Yazhuang Miao\textsuperscript{1},
Shuo Liang\textsuperscript{2}, Tianhui Qiu\textsuperscript{1},
and Libin Fu\textsuperscript{3}\\[5pt]
{\itshape
\textsuperscript{1}School of Science, Qingdao University of Technology, Qingdao, Shandong, China\\[2pt]
\textsuperscript{2}School of Electronic Engineering, Xidian University, Xi'an 710071, China\\[2pt]
\textsuperscript{3}Graduate School, China Academy of Engineering Physics, Beijing 100193, China
}
\end{center}
\vspace{6pt}

This Supplemental Material provides the derivations and numerical checks underlying the Letter.
We first derive the encoded and microwave-dressed NV Hamiltonians, distinguish
the effective pair interaction, its unequal-cell generalization, and the corrected pair resonance.
We then analyze the resonant one-phonon and off-resonant vacuum branches and derive the nonnegative-$P_{\rm GS}$
classical bound. The remaining sections specify the full charging sequence and battery
observables, establish the laboratory-frame energy balance to exhibit the contribution of energy
from phonon, define the operating-window scans, and examine the validity of the reduced
description on Hilbert space. Unless stated otherwise, all frequency-like quantities denote angular
frequencies.

We use the notation $\lvert n;xy\rangle\equiv\lvert n\rangle_m\otimes\lvert x_1y_2\rangle$
throughout this work, where $\lvert n\rangle_m$ is the $n$-phonon Fock state of the mechanical mode,
with $n=0,1,\ldots$. When $x,y\in\{g,e\}$, they denote microwave-dressed spin states; when $x,y\in\{0,1\}$,
they denote the encoded bare states.

\section{Microscopic NV model and dressed Hamiltonian}
\label{sec:S_microscopic}

\subsection{Laboratory-frame ground-manifold Hamiltonian}

Each negatively charged NV center has an electronic spin-triplet ground state with zero-field splitting $D/2\pi\simeq2.87~\mathrm{GHz}$~\cite{Doherty2013,PRL136200802}. We consider two centers with a common NV axis $z$
and encode the discharged state of cell $j$ as $\lvert0_j\rangle\equiv\lvert m_s=0\rangle_j$ and the
charged state as $\lvert1_j\rangle\equiv\lvert m_s=-1\rangle_j$. The dimensionless spin-1 operators obey
$S_j^z\lvert m_s\rangle=m_s\lvert m_s\rangle$. The encoded transitions in the two NV centers are
spectrally resolved and are addressed by separate microwave carriers. We work within a fixed nuclear-spin
manifold. Its static hyperfine contribution is therefore not retained as an explicit dynamical degree of
freedom but is included in the encoded transition frequency introduced below.

For fixed microwave amplitudes and carrier frequencies, the laboratory-frame Hamiltonian, from which the
encoded description is obtained, is
\begin{equation}
\begin{aligned}
\frac{H_{\rm lab}(t)}{\hbar}={}&\omega_m b^\dagger b
+\sum_{j=1}^{2}\left[D(S_j^z)^2+\gamma_eB_jS_j^z
+\gamma_eB_{{\rm mw},j}\cos(\omega_{d,j}t+\phi_j)S_j^x
+\lambda_{S,j}(b+b^\dagger)S_j^z\right]\\
&+d_{\rm dd}\left[\bm S_1\!\cdot\!\bm S_2
-3(\bm S_1\!\cdot\!\hat{\bm r}_{12})(\bm S_2\!\cdot\!\hat{\bm r}_{12})\right].
\end{aligned}
\label{eq:S_lab}
\end{equation}
Here $\omega_m$ is the mechanical-mode frequency and $b$ annihilates one phonon of that mode. $\gamma_e>0$
denotes the magnitude of the electron gyromagnetic ratio and $B_j$ the static magnetic field along the NV
axis at center $j$. For the microwave field of center $j$, $B_{{\rm mw},j}$ denotes the amplitude,
$\omega_{d,j}$ the carrier frequency, and $\phi_j$ the phase. $\bm S_j=(S_j^x,S_j^y,S_j^z)$ is the spin
vector. The axial spin-phonon coupling is $\lambda_{S,j}=\gamma_eG_jx_{\rm zpf}$, where $G_j$ is the local
axial field gradient and $x_{\rm zpf}$ is the mechanical zero-point displacement~\cite{Li2020}. The vector
$\bm r_{12}$ points from one NV center to the other, with $r_{12}=|\bm r_{12}|$ and
$\hat{\bm r}_{12}=\bm r_{12}/r_{12}$. Writing $\mu_0$ for the vacuum permeability,
$d_{\rm dd}=\mu_0\hbar\gamma_e^2/(4\pi r_{12}^3)$ sets the microscopic magnetic dipole--dipole scale.

\subsection{Projection onto the battery transition and microwave rotating frame}

Within the encoded two-level subspace, let
$\mathbb I_{j,{\rm enc}}=\lvert0_j\rangle\!\langle0_j\rvert+\lvert1_j\rangle\!\langle1_j\rvert$ denote the
identity, and define $\tau_j^z=\lvert1_j\rangle\!\langle1_j\rvert-\lvert0_j\rangle\!\langle0_j\rvert$ and
$\tau_j^x=\lvert0_j\rangle\!\langle1_j\rvert+\lvert1_j\rangle\!\langle0_j\rvert$. The spin-1 operators
restricted to this subspace satisfy  $S_j^z\mapsto-(\mathbb I_{j,{\rm enc}}+\tau_j^z)/2$,
$(S_j^z)^2\mapsto(\mathbb I_{j,{\rm enc}}+\tau_j^z)/2$, and $S_j^x\mapsto\tau_j^x/\sqrt2$.

We denote the actual $\lvert0_j\rangle\leftrightarrow\lvert1_j\rangle$ transition frequency, including the
fixed hyperfine contribution from nuclear-spin, by $\omega_{{\rm enc},j}$; its electronic contribution from
Eq.~\eqref{eq:S_lab} is $D-\gamma_eB_j$. Before the static shifts introduced below are included, the microwave
detuning is $\delta_{j,\rm raw}=\omega_{{\rm enc},j}-\omega_{d,j}$. Choosing each carrier phase so that the
drive lies along the encoded $x$ axis, the transformation to the corresponding carrier frame followed by the
microwave rotating-wave approximation gives $(\delta_{j,\rm raw}\tau_j^z+\Omega_j\tau_j^x)/2$, with
$\Omega_j=\gamma_eB_{{\rm mw},j}/\sqrt2$.

Projection of the magnetic-gradient term gives $\lambda_{S,j}(b+b^\dagger)S_j^z
\mapsto \lambda_j(b+b^\dagger)(\mathbb I_{j,{\rm enc}}+\tau_j^z)$,
where $\lambda_j=-\lambda_{S,j}/2$. The identity component produces a spin-independent force
$(\lambda_1+\lambda_2)(b+b^\dagger)$. The displacement $b=\tilde b-(\lambda_1+\lambda_2)/\omega_m$
removes this force and generates the local term $-2(\lambda_1+\lambda_2)\lambda_j\tau_j^z/\omega_m$.
Accordingly, the detuning after the displacement is
$\delta_{j,\rm disp} =\delta_{j,\rm raw}-4(\lambda_1+\lambda_2)\lambda_j/\omega_m$.
For $\lambda_1=\lambda_2=\lambda$, the displacement contribution is $-8\lambda^2/\omega_m$.
We subsequently relabel the displaced oscillator $\tilde b$ as $b$.
\subsection{Dipolar interaction and carrier-secular reduction}

Let $\beta$ denote the angle between the inter-center axis $\hat{\bm r}_{12}$ and the common NV
axis. The part of the dipolar interaction remains the longitudinal term in the local spin projections
as $d_{\rm dd}(1-3\cos^2\beta)S_1^zS_2^z$. The transverse dipolar terms contain exchange processes
oscillating at the difference of the two addressed transition frequencies and double-raising or
double-lowering processes oscillating at their sum. For spectrally resolved centers satisfying
$d_{\rm dd}\ll|\omega_{{\rm enc},1}-\omega_{{\rm enc},2}|$, these terms are removed by the
individual-carrier rotating-wave (carrier-secular) reduction.

Projecting the surviving longitudinal term onto the encoded subspace gives
$d_{\rm dd}(1-3\cos^2\beta)S_1^zS_2^z\mapsto
J(\mathbb I_{1,{\rm enc}}\mathbb I_{2,{\rm enc}}+\tau_1^z+\tau_2^z+\tau_1^z\tau_2^z)$,
where $J=d_{\rm dd}(1-3\cos^2\beta)/4$. The first term is an irrelevant constant, while the two local
terms shift the encoded detunings by $2J$. We therefore define the detunings used from this point
onward as $\delta_j=\delta_{j,\rm disp}+2J$. The remaining term $J\tau_1^z\tau_2^z$ is the nonlocal
dipolar interaction. For the transverse geometry used throughout this work, $\beta=\pi/2$ and hence
$J=d_{\rm dd}/4$.

Combining the projected microwave, magnetic-gradient, and dipolar terms yields
\begin{equation}
H_{\rm enc}/\hbar
=\omega_mb^\dagger b
+\sum_j(\delta_j\tau_j^z+\Omega_j\tau_j^x)/2
+\sum_j\lambda_j(b+b^\dagger)\tau_j^z
+J\tau_1^z\tau_2^z,
\label{Hencc}
\end{equation}
which is the encoded Hamiltonian used immediately before the microwave-dressed form in
Eq.~\eqref{eq:centralH} in the main text.

\subsection{Microwave dressing and interaction vertices}
\label{sec:S_dressing}

The local driven-spin Hamiltonian (Eq. (\ref{Hencc}) excludes $J\tau_1^z\tau_2^z$) is diagonalized
by a rotation about the $y$ axis. We define
$\Lambda_j=(\delta_j^2+\Omega_j^2)^{1/2}$,
$\cos\theta_j=\delta_j/\Lambda_j$, and
$\sin\theta_j=\Omega_j/\Lambda_j$.
Let
$\{\lvert g_j\rangle,\lvert e_j\rangle\}$
denote the eigenstates of the local driven Hamiltonian, with the dressed Pauli operator defined by
$\sigma_j^z\lvert g_j\rangle=-\lvert g_j\rangle$ and $\sigma_j^z\lvert e_j\rangle=\lvert e_j\rangle$.
In this basis,
$\tau_j^z=\cos\theta_j\,\sigma_j^z-\sin\theta_j\,\sigma_j^x$,
$\tau_j^x=\sin\theta_j\,\sigma_j^z+\cos\theta_j\,\sigma_j^x$,
which gives Eq.~\eqref{eq:centralH} in the main text.

For the symmetric case used to expose the pair mechanism analytically, we set
$\Lambda_1=\Lambda_2=\Lambda$,
$\theta_1=\theta_2=\theta$, and
$\lambda_1=\lambda_2=\lambda$.
Writing $c=\cos\theta$ and $s=\sin\theta$, the interaction part of Eq.~\eqref{eq:centralH} in the main
text separates as
\begin{equation}
\begin{aligned}
&\frac{V_\lambda}{\hbar}=\sum_j\lambda_j(b+b^\dagger)\tau_j^z={}\lambda(b+b^\dagger)
\left[c(\sigma_1^z+\sigma_2^z)-s(\sigma_1^x+\sigma_2^x)\right],\\
&\frac{V_J}{\hbar}=J\tau_1^z\tau_2^z={}J\left[c^2\sigma_1^z\sigma_2^z
-cs(\sigma_1^z\sigma_2^x+\sigma_1^x\sigma_2^z)
+s^2\sigma_1^x\sigma_2^x\right].
\end{aligned}
\label{eq:S_vertices}
\end{equation}
We use $V\equiv V_\lambda+V_J$ for the total interaction in the perturbative pair analysis below.

\section{Effective pair interaction and corrected resonance}
\label{sec:S_SW}

\subsection{Pair coupling}

For $n\ge1$, we denote the two states forming the pair states by $\lvert i_n\rangle\equiv\lvert n;gg\rangle$
and $\lvert f_n\rangle\equiv\lvert n-1;ee\rangle$. Under $H_0/\hbar=\omega_m b^\dagger b +\Lambda(\sigma_1^z+\sigma_2^z)/2$,
their unperturbed energies are $E_{i_n}/\hbar=n\omega_m-\Lambda$ and $E_{f_n}/\hbar=(n-1)\omega_m+\Lambda$,
so their unperturbed separation is $(E_{f_n}-E_{i_n})/\hbar=2\Lambda-\omega_m$. They are therefore degenerate
at the unperturbed pair resonance $\omega_m=2\Lambda$.

The spin--phonon interaction $V_\lambda$ changes the phonon number by one and can change at most one dressed spin,
whereas the dipolar interaction $V_J$ changes the dressed-spin configuration without changing the phonon number.
Consequently, $\langle f_n\rvert V\lvert i_n\rangle=0$, no single term in $V$ (Eq. (\ref{eq:S_vertices})) directly
connects the two pair states. However, the second-order virtual paths jointly generated by the two interactions
connecting the pair states.

In quasi-degenerate perturbation situation, let $H_0\lvert\mu\rangle=E_\mu\lvert\mu\rangle$. The second-order
effective Hamiltonian $H_{\rm eff}^{(2)}$ has the off-diagonal matrix element
\begin{equation}
\begin{aligned}
\langle f_n\rvert H_{\rm eff}^{(2)}\lvert i_n\rangle
=
\frac{1}{2}
\sum_{\mu}^{\prime}
\langle f_n\rvert V\lvert\mu\rangle
\langle\mu\rvert V\lvert i_n\rangle
\left(
\frac{1}{E_{i_n}-E_\mu}
+
\frac{1}{E_{f_n}-E_\mu}
\right),
\end{aligned}
\label{eq:S_SW_general}
\end{equation}
where the prime excludes the two pair states from the sum.

At the unperturbed degeneracy, $E_{i_n}=E_{f_n}\equiv E_{\rm deg}$, only cross terms containing one
$V_\lambda$ vertex and one $V_J$ vertex connect the two endpoints. They give the six nonzero
virtual contributions listed in Table~\ref{tab:S_paths}. Four paths pass through states with one
dressed spin excited and have energy denominator $\pm\hbar\Lambda$; the remaining two pass through
$\lvert n-1;gg\rangle$ or $\lvert n;ee\rangle$ have denominator $\pm2\hbar\Lambda$.

\begin{table}[t]
\caption{
Second-order $J\lambda$ virtual paths connecting
$\lvert i_n\rangle=\lvert n;gg\rangle$ and
$\lvert f_n\rangle=\lvert n-1;ee\rangle$ at the unperturbed pair degeneracy
$\omega_m=2\Lambda$.
The two matrix-element columns are divided by $\hbar$.
The ``first vertex'' column specifies the interaction acting on
$\lvert i_n\rangle$.
The final column gives each contribution to $g_n$ in units of
$J\lambda cs^2\sqrt n/\Lambda$.
}
\label{tab:S_paths}
\centering
\small
\begin{tabular}{cccccc}
\toprule
Intermediate state
& first vertex
& $\langle\mu|V|i_n\rangle/\hbar$
& $\langle f_n|V|\mu\rangle/\hbar$
& $(E_{\rm deg}-E_\mu)/\hbar$
& relative contribution \\
\midrule
$\lvert n-1;eg\rangle$
& $V_\lambda$
& $-\lambda s\sqrt n$
& $-Jcs$
& $+\Lambda$
& $+1$ \\

$\lvert n-1;ge\rangle$
& $V_\lambda$
& $-\lambda s\sqrt n$
& $-Jcs$
& $+\Lambda$
& $+1$ \\

$\lvert n-1;gg\rangle$
& $V_\lambda$
& $-2\lambda c\sqrt n$
& $Js^2$
& $+2\Lambda$
& $-1$ \\

$\lvert n;eg\rangle$
& $V_J$
& $Jcs$
& $-\lambda s\sqrt n$
& $-\Lambda$
& $+1$ \\

$\lvert n;ge\rangle$
& $V_J$
& $Jcs$
& $-\lambda s\sqrt n$
& $-\Lambda$
& $+1$ \\

$\lvert n;ee\rangle$
& $V_J$
& $Js^2$
& $2\lambda c\sqrt n$
& $-2\Lambda$
& $-1$ \\
\bottomrule
\end{tabular}
\end{table}

The four paths through the single-spin manifolds add constructively. The remaining two paths
involving the longitudinal spin--phonon vertex and the double-transverse dipolar vertex have
the opposite sign and partially cancel them. The coherent sum gives the off-diagonal pair-coupling in
Eq.~\eqref{eq:pairH}
\begin{equation}
g_n(\Lambda,\theta)
\equiv
\frac{\langle f_n\rvert H_{\rm eff}^{(2)}
\lvert i_n\rangle}{\hbar}
=
\sqrt n\,g_{\rm pair}(\Lambda,\theta),
\qquad
g_{\rm pair}(\Lambda,\theta)
=
2\frac{J\lambda}{\Lambda}\cos\theta\sin^2\theta\,
\label{eq:S_gn}
\end{equation}
The bilinear dependence $g_n\propto J\lambda$ reflects the necessity for both interactions in (\ref{eq:S_vertices}),
whereas the factor $\cos\theta\sin^2\theta$ reflects the simultaneous involvement of longitudinal and transverse
dressed components. Within $0\le\theta\le\pi/2$, the pair-coupling coefficient therefore vanishes at $\theta=0$ and $\theta=\pi/2$.

The angular factor $\cos\theta\sin^2\theta$ is maximal for $\cos^2\theta=1/3$ within $0\le\theta\le\pi/2$.
We therefore use $\theta_0=\arccos(1/\sqrt3)$ as the reference mixing angle. At this reference point,
$g_{\rm p}=|g_{\rm pair}(\Lambda_0,\theta_0)|$ and the leading one-phonon resonant transfer time is
$t_\pi=\pi/(2g_{\rm p})$. At the reference dressed controls, a higher Fock sector has the corresponding effective
transfer time $t_\pi/\sqrt n$ when tuned to its pair resonance. The vacuum sector has no corresponding phonon-annihilation
endpoint and is treated separately in Sec.~\ref{sec:S_vacuum}.

\subsection{Unequal-cell pair coupling}
\label{sec:S_unequal}

For general dressed splittings $\Lambda_j$, mixing angles $\theta_j$, and spin--phonon couplings $\lambda_j$,
we write $c_j=\cos\theta_j$ and $s_j=\sin\theta_j$. The two pair states $\lvert i_n\rangle=\lvert n;gg\rangle$
and $\lvert f_n\rangle=\lvert n-1;ee\rangle$ have the unperturbed separation $\hbar(\Lambda_1+\Lambda_2-\omega_m)$,
so the leading resonance condition becomes $\omega_m=\Lambda_1+\Lambda_2$. Evaluating the same virtual paths
at this degeneracy gives
\begin{equation}
g_n
=
2\sqrt n\,J s_1s_2
\left[
\frac{\lambda_1c_1}{\Lambda_2}
+\frac{\lambda_2c_2}{\Lambda_1}
-\frac{\lambda_1c_1+\lambda_2c_2}
{\Lambda_1+\Lambda_2}
\right].
\label{eq:S_generalUnequalPair}
\end{equation}
For
$\Lambda_1=\Lambda_2=\Lambda$,
$\theta_1=\theta_2=\theta$, and
$\lambda_1=\lambda_2=\lambda$,
Eq.~\eqref{eq:S_generalUnequalPair} reduces directly to
Eq.~\eqref{eq:S_gn}.

To isolate the effect of a differential dressed-splitting mismatch,
we take
$\Lambda_{1,2}=\bar\Lambda\pm\Delta_\Lambda$
while keeping
$\theta_1=\theta_2=\theta$ and
$\lambda_1=\lambda_2=\lambda$.
Equation~\eqref{eq:S_generalUnequalPair} then gives
\begin{equation}
\frac{g_n(\Delta_\Lambda)}
     {g_n(0)}
=
\frac{\bar\Lambda^2+\Delta_\Lambda^2}
     {\bar\Lambda^2-\Delta_\Lambda^2}
=
1+
2\left(\frac{\Delta_\Lambda}{\bar\Lambda}\right)^2
+
O\!\left[
\left(\frac{\Delta_\Lambda}{\bar\Lambda}\right)^4
\right].
\label{eq:S_splitAsymmetry}
\end{equation}
The leading pair coupling therefore has no linear correction in an antisymmetric splitting
mismatch about the symmetric point.

Likewise, for
$\Lambda_1=\Lambda_2=\Lambda$ and
$\theta_1=\theta_2=\theta$, let
$\lambda_{1,2}=\bar\lambda\pm\Delta_\lambda$
at fixed mean coupling $\bar\lambda$.
Equation~\eqref{eq:S_generalUnequalPair} becomes
\begin{equation}
g_n
=
2\sqrt n\,\cos\theta\sin^2\theta\,
\frac{J\bar\lambda}{\Lambda},
\label{eq:S_lambdaAsymmetry}
\end{equation}
so the leading $J\lambda$ pair coupling is independent of the
antisymmetric coupling difference $\Delta_\lambda$.

These results concern the off-diagonal pair coupling itself.
A mismatch can still shift the pair resonance and, for a control
sequence of fixed duration, can also change the observed transfer
through detuning or pulse-area errors. The diagonal corrections that
set the resonance position are considered next.

\subsection{Corrected pair resonance}
\label{sec:S_shift}

The unperturbed condition $\omega_m=2\Lambda$ does not coincide
exactly with the avoided-crossing center because the interactions shift
the two pair states differently. The resonance displacement has no
first-order contribution: the diagonal dipolar term
$Jc^2\sigma_1^z\sigma_2^z$ gives the same shift to both pair states,
whereas the first-order spin--phonon expectation value vanishes because
${}_m\!\langle n|b+b^\dagger|n\rangle_m=0$.
We denote the total second-order frequency shift of
$\lvert n;gg\rangle$ by $\Sigma_{n,gg}^{(2)}$ and that of
$\lvert n-1;ee\rangle$ by $\Sigma_{n-1,ee}^{(2)}$. We separate each
shift into dipolar and spin--phonon contributions.

The off-diagonal dipolar vertices give
\begin{equation}
\begin{aligned}
\Sigma_{n,gg;J}^{(2)}
&=
-\frac{2J^2c^2s^2}{\Lambda}
-\frac{J^2s^4}{2\Lambda},\\
\Sigma_{n-1,ee;J}^{(2)}
&=
+\frac{2J^2c^2s^2}{\Lambda}
+\frac{J^2s^4}{2\Lambda},
\end{aligned}
\qquad
\Sigma_{n-1,ee;J}^{(2)}-\Sigma_{n,gg;J}^{(2)}
=
\frac{J^2}{\Lambda}
\left(4c^2s^2+s^4\right).
\label{eq:S_Jshift}
\end{equation}
Equation~\eqref{eq:S_Jshift} contains the complete $J^2$ differential shift. Cross terms proportional to $J\lambda$ do not contribute to these
diagonal shifts: $V_J$ preserves the phonon number, whereas a single
$V_\lambda$ vertex changes it by one, so one vertex of each type cannot
return a pair state to itself.

For the spin--phonon interaction, the longitudinal dressed component
produces the same second-order shift $-4\lambda^2c^2/\omega_m$ for both
pair states and therefore cancels from their difference. Including the
transverse virtual processes gives
\begin{equation}
\begin{aligned}
\Sigma_{n,gg;\lambda}^{(2)}
={}&
-\frac{4\lambda^2c^2}{\omega_m}
+
2\lambda^2s^2
\left[
\frac{n}{\omega_m-\Lambda}
-\frac{n+1}{\omega_m+\Lambda}
\right],\\
\Sigma_{n-1,ee;\lambda}^{(2)}
={}&
-\frac{4\lambda^2c^2}{\omega_m}
+
2\lambda^2s^2
\left[
\frac{n-1}{\omega_m+\Lambda}
-\frac{n}{\omega_m-\Lambda}
\right].
\end{aligned}
\label{eq:S_lambdashifts}
\end{equation}
The common term in Eq.~\eqref{eq:S_lambdashifts} is a dressed-state
second-order spin--phonon shift and is distinct from the
spin-independent static displacement already absorbed into the encoded
detunings in Sec.~\ref{sec:S_microscopic}. The differential
spin--phonon shift is
$\Sigma_{n-1,ee;\lambda}^{(2)}-\Sigma_{n,gg;\lambda}^{(2)}
=-8n\lambda^2s^2\Lambda/(\omega_m^2-\Lambda^2)$.
Thus
$\Sigma_{n,gg}^{(2)}=\Sigma_{n,gg;J}^{(2)}+\Sigma_{n,gg;\lambda}^{(2)}$
and
$\Sigma_{n-1,ee}^{(2)}=\Sigma_{n-1,ee;J}^{(2)}+\Sigma_{n-1,ee;\lambda}^{(2)}$,
consistent with the shifts used in Eq.~\eqref{eq:pairH} of the Letter.

Combining the unperturbed separation with the differential second-order
shifts yields
\begin{equation}
\delta_{{\rm p},n}
=
2\Lambda-\omega_m
+
\frac{J^2}{\Lambda}
\left(4c^2s^2+s^4\right)
-
\frac{8n\lambda^2s^2\Lambda}
{\omega_m^2-\Lambda^2}
+
O\!\left[\max(|J|,|\lambda|)^3/\Lambda^2\right].
\label{eq:S_pairdetuning}
\end{equation}
Expanding the resonance condition in Eq.~\eqref{eq:S_pairdetuning}
about the unperturbed pair resonance $\omega_m=2\Lambda$ and retaining
terms consistently through second order gives
\begin{equation}
\frac{\omega_{{\rm res},n}^{(2)}}{\Lambda}
\simeq 2 + \left(\frac{J}{\Lambda}\right)^2 \left(4c^2s^2+s^4\right) - \frac{8n}{3} \left(\frac{\lambda}{\Lambda}\right)^2s^2 .
\label{eq:S_resonance}
\end{equation}
The last term makes the second-order resonance weakly Fock-sector
dependent. We therefore use $\omega_{{\rm res},n}^{(2)}$ for the
analytic second-order resonance of sector $n$ and reserve the
projected-model crossing center for the value obtained by direct diagonalization of
Eq.~\eqref{eq:centralH}. In the Fock-sector scaling comparison, each
projected-model trajectory is evaluated at its own crossing center so that the
small sector-dependent resonance shift does not enter the comparison of
$g_n$.

At the reference point,
$\omega_{\rm res}^{(2)}\equiv
\omega_{{\rm res},1}^{(2)}(\Lambda_0,\theta_0)$.
For the parameters used here,
$\omega_{\rm res}^{(2)}/\Lambda_0\simeq2.007513$, whereas direct
diagonalization gives $\omega_{\rm res}/\Lambda_0\simeq2.007484$.
Their difference is much smaller than $g_{\rm p}$ and measures the
higher-order correction to the second-order resonance position within
the projected model.

\section{Pair-transfer benchmarks and classical-input bound}
\label{sec:S_benchmarks}

\subsection{Vacuum branch and off-resonant hierarchy}
\label{sec:S_vacuum}

For the mechanical vacuum, the lower-phonon partner state required by the resonant pair transition
does not exist. Starting from $\lvert0;gg\rangle$, every state with both dressed spins excited has
the form $\lvert\ell;ee\rangle$ with $\ell\ge0$, and its unperturbed separation from the initial
state is $(E_{\ell,ee}-E_{0,gg})/\hbar=\ell\omega_m+2\Lambda$. This separation remains at least
$2\Lambda$, which means the vacuum sector therefore contains no quasi-degenerate doubly excited
partner state near the pair resonance.

The leading vacuum background is generated directly by the double-transverse dipolar term
$Js^2\sigma_1^x\sigma_2^x$, with $\lvert0;gg\rangle\leftrightarrow\lvert0;ee\rangle$ matrix element
being $Js^2$ and the detuning being $2\Lambda$. To leading order in $J/\Lambda$, the contribution
of this channel to the double-excitation population is
\begin{equation}
P_{ee,{\rm vac};J}(t)
=
s^4\left(\frac{J}{\Lambda}\right)^2
\sin^2(\Lambda t),
\label{eq:S_vacDirect}
\end{equation}
so Eq.~\eqref{eq:S_vacDirect} is bounded by $s^4(J/\Lambda)^2$ independently of the resonant pair-transfer time.
The $J\lambda$ processes that generate the resonant one-phonon transition also have vacuum counterparts, but they
are counterrotating: because no phonon can be annihilated from the vacuum, they connect $\lvert0;gg\rangle$ to
$\lvert1;ee\rangle$, whose detuning is $\omega_m+2\Lambda\simeq4\Lambda$. Their contribution to the probability
is therefore of order $J^2\lambda^2/\Lambda^4$. Processes involving two spin--phonon vertices are likewise
off-resonant and enter only at higher order in $\lambda/\Lambda$. Consequently, the vacuum double-excitation
population is dominated in the weak-coupling regime by the dipolar background, $P_{ee,{\rm vac}}(t)=O[(J/\Lambda)^2]$.

The one-phonon branch contains a quasi-degenerate two-state manifold and must be treated by the effective pair
Hamiltonian of Eq.~\eqref{eq:pairH}, whereas the vacuum branch contains only off-resonant transitions and remains
perturbative. The distinction between the two branches is therefore structural. The vacuum response shown in
Fig.~\ref{fig:gating}(d) is obtained from direct numerical simulation of Eq.~\eqref{eq:centralH}, which retains all
of these finite-coupling backgrounds.

\subsection{Finite-coupling corrections on the one-phonon branch}
\label{sec:S_onbranch}

For $n=1$, the pair subspace is spanned by $\lvert i_1\rangle=\lvert1;gg\rangle$ and $\lvert f_1\rangle=\lvert0;ee\rangle$,
as defined in Sec.~\ref{sec:S_SW}. Beyond the leading second-order reduction above, virtual processes renormalize both the
pair detuning and the off-diagonal coupling. The differences $|\omega_{\rm res}-\omega_{\rm res}^{(2)}|$ and $|t_{\star,1}-t_\pi|$
directly quantify these corrections to the resonance position and transfer time. Admixture outside the pair subspace sets
the scale of the residual deviation from ideal two-state transfer.

The leading off-subspace admixture weights of the two pair states are
\begin{equation}
\chi_i
=
\sum_{\mu}^{\prime}
\frac{
|\langle\mu|V|i_1\rangle|^2
}{
(E_{i_1}-E_\mu)^2
},
\qquad
\chi_f
=
\sum_{\mu}^{\prime}
\frac{
|\langle\mu|V|f_1\rangle|^2
}{
(E_{f_1}-E_\mu)^2
}.
\label{eq:S_endpointAdmixture}
\end{equation}
Here $|\mu\rangle$ denotes the off-subspace eigenstates of $H_0/\hbar=\omega_m b^\dagger b +\Lambda(\sigma_1^z+\sigma_2^z)/2$,
and the prime excludes the two pair states from the sums. Evaluating Eq.~\eqref{eq:S_endpointAdmixture} at the
unperturbed pair point $\omega_m=2\Lambda$ gives
\begin{equation}
\chi_i=
\left(3c^2+\frac{22}{9}s^2\right)
\left(\frac{\lambda}{\Lambda}\right)^2
+
\left(2c^2s^2+\frac{s^4}{4}\right)
\left(\frac{J}{\Lambda}\right)^2,
\label{eq:S_chiI}
\end{equation}
and
\begin{equation}
\chi_f=
\left(c^2+2s^2\right)
\left(\frac{\lambda}{\Lambda}\right)^2
+
\left(2c^2s^2+\frac{s^4}{4}\right)
\left(\frac{J}{\Lambda}\right)^2.
\label{eq:S_chiF}
\end{equation}
Using the corrected rather than the unperturbed pair resonance in these denominators changes
Eqs.~\eqref{eq:S_chiI} and \eqref{eq:S_chiF} only at higher perturbative order.

The quantities $\chi_i$ and $\chi_f$ characterize the leading perturbative dressing of the
two pair states; they are not themselves dynamical leakage probabilities. When the
one-phonon dynamics is evaluated at its projected avoided-crossing center and first maximum,
their sum sets the scale of the remaining finite-coupling infidelity,
$1-P_{ee}^{\max}=O(\chi_i+\chi_f)$, where $P_{ee}^{\max}$ denotes the peak dressed double-excitation
population. At the reference angle $c^2=1/3$,
\begin{equation}
\chi_i+\chi_f
=
\frac{116}{27}
\left(\frac{\lambda}{\Lambda}\right)^2
+
\frac{10}{9}
\left(\frac{J}{\Lambda}\right)^2,
\label{eq:S_totalAdmixture}
\end{equation}
consistent with the residual deviation from unit transfer in the projected dynamics.

Equation~\eqref{eq:S_totalAdmixture} also makes the quality--speed tradeoff. If both
physical couplings are reduced as $J\rightarrow\kappa J$ and $\lambda\rightarrow\kappa\lambda$
at fixed ratios and fixed $\Lambda$, the vacuum background and the projected one-phonon
finite-coupling error decrease as $O(\kappa^2)$. At the same time,
$g_{\rm pair}\propto J\lambda/\Lambda=O(\kappa^2)$, so the resonant transfer time scales
as $t_\pi=O(\kappa^{-2})$. Increasing the separation between the on- and off-resonant
responses therefore comes at the cost of a slower second-order pair transfer.

\subsection{Nonnegative-$P_{\rm GS}$ classical bound}
\label{sec:S_classical}

The classical-input bound used in this Letter applies to any fixed configuration with a
normalized response. The initial spin state, Hamiltonian, control waveform, and readout
time are held fixed, while only the initial mechanical state is varied. Such a response
can be written as
\begin{equation}
R(M,\rho_m)=\operatorname{Tr}_m(M\rho_m),
\qquad
0\le M\le\mathbb I_m,
\label{eq:S_genericEffect}
\end{equation}
where $\rho_m$ is the initial mechanical density operator, $M$ is an effect operator on the
initial mechanical Hilbert space, $\mathbb I_m$ is the identity on that space, and
$R_{\rm vac}=R(M,|0\rangle\langle 0|)={}_m\!\langle0|M|0\rangle_m$ is the response to the pure
mechanical vacuum. For the pair-transfer benchmark at time $t$, let $U(t)$ denote the joint
spin--mechanical propagator generated at the fixed configuration. Then to evaluate the effect
from $|gg\rangle$ to $|ee\rangle$, equation~\eqref{eq:S_genericEffect} uses the effect operator
$M_{ee}(t)\equiv\langle gg|U^\dagger(t)[\mathbb I_m\otimes|ee\rangle\langle ee|]U(t)|gg\rangle$,
so that $R(M_{ee}(t),\rho_m)=P_{ee}(t)$. At finite coupling, $M_{ee}(t)$ need not be diagonal
in the phonon number basis; therefore the bound allows phonon-number coherences.

A classical mechanical state, including coherent state, has a nonnegative Glauber--Sudarshan
representation $\rho_{\rm cl}=\int d^2\alpha\,P_{\rm GS}(\alpha) |\alpha\rangle_m{}_m\!\langle\alpha|$,
with $P_{\rm GS}(\alpha)\ge0$ and $\int d^2\alpha\,P_{\rm GS}(\alpha)=1$~\cite{Glauber1963,Sudarshan1963}.
For fixed mean occupation $\bar n=\int d^2\alpha\,P_{\rm GS}(\alpha)|\alpha|^2$, its vacuum
weight obeys
\begin{equation}
p_0
\equiv
{}_m\!\langle0|\rho_{\rm cl}|0\rangle_m
=
\int d^2\alpha\,P_{\rm GS}(\alpha)e^{-|\alpha|^2}
\ge e^{-\bar n},
\label{eq:S_vacuumweight}
\end{equation}
where the inequality follows from Jensen's inequality for the convex function $e^{-x}$.
Equation~\eqref{eq:S_vacuumweight} is the classical-input constraint used below.

Let $R_{\rm cl}\equiv R(M_{ee}(t),\rho_{\rm cl})$. Using the trace norm $\|\cdot\|_1$, we define
the root fidelity as $F(\rho_1,\rho_2)=\|\sqrt{\rho_1}\sqrt{\rho_2}\|_1$. The root fidelity
between the pure vacuum and $\rho_{\rm cl}$ is $\sqrt{p_0}$ which would be preserved under the
unitary evolution $U(t)$ with the fixed configuration, and the two-outcome measurement
$\{M,\mathbb I_m-M\}$ cannot decrease $F(\rho_1,\rho_2)$. Hence
\begin{equation}
\sqrt{p_0}
\le
\sqrt{R_{\rm cl}R_{\rm vac}}
+
\sqrt{(1-R_{\rm cl})(1-R_{\rm vac})}.
\label{eq:S_fidelity}
\end{equation}
In the regime relevant here, $R_{\rm vac}<e^{-\bar n}$, so every
allowed classical state has $p_0\ge e^{-\bar n}>R_{\rm vac}$. Solving
Eq.~\eqref{eq:S_fidelity} for the largest allowed response gives
\begin{equation}
R_{\rm cl}
\le
\left[
\sqrt{p_0R_{\rm vac}}
+
\sqrt{(1-p_0)(1-R_{\rm vac})}
\right]^2 .
\label{eq:S_overlapbound}
\end{equation}
For fixed $R_{\rm vac}$, the right-hand side of
Eq.~\eqref{eq:S_overlapbound} decreases with $p_0$ increasing over this interval. The least restrictive classical bound therefore occurs
at the minimum allowed vacuum weight, $p_0=e^{-\bar n}$:
\begin{equation}
R_{\rm cl}
\le
\mathcal C_{\bar n}(R_{\rm vac})
\equiv
\left[
\sqrt{e^{-\bar n}R_{\rm vac}}
+
\sqrt{(1-e^{-\bar n})(1-R_{\rm vac})}
\right]^2 .
\label{eq:S_generalClassicalBound}
\end{equation}

For the pair-transfer benchmark, we set $\bar n=1$,
$\omega_m=\omega_{\rm res}^{(2)}$, and the common readout time to
$t_\pi$. Substituting $R_{\rm vac}=P_{ee,{\rm vac}}(t_\pi)$ into
Eq.~\eqref{eq:S_generalClassicalBound} gives
Eq.~\eqref{eq:classicalBound} of this Letter. In the ideal vacuum-off
limit $P_{ee,{\rm vac}}(t_\pi)\rightarrow0$, the bound becomes
$1-e^{-1}$. Direct propagation of Eq.~\eqref{eq:centralH} at this
operation gives $t_\pi=0.714879~\mathrm{ms}$ and
$P_{ee,{\rm vac}}(t_\pi)=1.10059\times10^{-3}$, corresponding to a
classical upper bound of $0.66381$, while the one-phonon response is
$P_{ee}(t_\pi)=0.99140$. These three values are compared in
Fig.~\ref{fig:gating}(d).

\section{Full charging sequence and battery observables}
\label{sec:S_fullcycle}

\subsection{Controlled charging sequence}
\label{sec:S_freeze}
During this charging process, the fixed-carrier microwave term in Eq.~\eqref{eq:S_lab}
is generalized to
$\gamma_eB_{{\rm mw},j}(t)
\cos[\Phi_{d,j}(t)]S_j^x$, with
\[
\Phi_{d,j}(t)
=
\phi_j+\int_0^t dt'\,\omega_{d,j}(t').
\]
The corresponding time-dependent carrier transformation followed by
the microwave rotating-wave approximation gives
$\Omega_j(t)=\gamma_eB_{{\rm mw},j}(t)/\sqrt2$ and
\[
\delta_j(t)
=
\omega_{{\rm enc},j}-\omega_{d,j}(t)
-\frac{4(\lambda_1+\lambda_2)\lambda_j}{\omega_m}
+2J .
\]
For the symmetric protocol, the two microwave drives have the same effective detuning and Rabi
frequency, $\delta_1(t)=\delta_2(t)=\delta(t)$ and $\Omega_1(t)=\Omega_2(t)=\Omega(t)$. The Rabi
frequency is controlled by the microwave amplitude, while the detuning is controlled by the
carrier frequency at fixed static fields $B_j$. The common detuning modulation shifts both carrier
frequencies by the same amount and preserves their spectral separation. The five-stage sequence
is described by the time-dependent encoded Hamiltonian
\begin{equation}
\frac{H_{\rm enc}(t)}{\hbar}
=
\omega_m b^\dagger b
+\sum_{j=1}^{2}
\left[
\frac{\delta(t)}{2}\tau_j^z
+\frac{\Omega(t)}{2}\tau_j^x
+\lambda(b+b^\dagger)\tau_j^z
\right]
+J\tau_1^z\tau_2^z ,
\label{eq:S_controlH}
\end{equation}
where $\delta(t)=\Lambda(t)\cos\theta(t)$, $\Omega(t)=\Lambda(t)\sin\theta(t)$, and
$\Lambda(t)=(\delta(t)^2+\Omega(t)^2)^{1/2}$.

We write $\rho(t)$ as the state for the joint spin--mechanical system. The two cells are driven
symmetrically throughout the sequence, and the mechanical frequency is fixed at
$\omega_m=\omega_{\rm res}^{(2)}$. For each finite-duration ramp we use the local time-scaling
of a charging stage $u\in[0,1]$ and the cubic smoothstep $\mathcal R(u)=3u^2-2u^3$, which
satisfies $\mathcal R(0)=0$, $\mathcal R(1)=1$, and $\mathcal R'(0)=\mathcal R'(1)=0$. The
preparation and off-resonant splittings are $\Lambda_{\rm prep}=1.10\Lambda_0$ and
$\Lambda_{\rm off}=\Lambda_0+20g_{\rm p}$.

During \emph{dress}, we keep $\Lambda=\Lambda_{\rm prep}$ and vary $\theta(u)=\theta_0\mathcal R(u)$.
Starting from $\lvert1;00\rangle$, the local spins follow the instantaneous dressed ground states
toward $\lvert1;gg\rangle$ while the pair transition remains off resonance. We use
$\tau_{\rm dress}=100/\Lambda_0\simeq22.74~\mu{\rm s}$. For the reference parameters in the main text,
the single-phonon detuning-to-coupling ratio $\delta_{{\rm p},1}/g_{\rm pair}$ is smallest at $\theta_0$ and
equals $4.39\times10^2$, while the smoothstep gives the maximum local nonadiabatic amplitude of order
$\max|\dot\theta|/\Lambda_{\rm prep}=1.30\times10^{-2}$, consistent with local adiabatic following.
For the one-phonon input, the dressed-ground-state population at the end of the ramp is $P_{gg}=0.99704$.
These numerical results support the \emph{dress} stage in Fig.~3(b) of the main text.

During \emph{tune}, $\theta=\theta_0$ is fixed and $\Lambda(u)=\Lambda_{\rm prep}
+(\Lambda_0-\Lambda_{\rm prep})\mathcal R(u)$.
Because the local dressed eigenstates depend on $\theta$, resulting from the form of the local
Hamiltonian, but not on the magnitude of $\Lambda$, this ramp changes the pair detuning without
changing the local dressed basis. We use $\tau_{\rm tune}=0.01t_\pi\simeq7.149~\mu{\rm s}$, for
which $g_{\rm p}\tau_{\rm tune}=\pi/200\simeq1.57\times10^{-2}$. The total dressed double-excitation
population at transfer-stage entry is $P_{ee}=7.63\times10^{-4}$.

During \emph{transfer}, $(\Lambda,\theta)=(\Lambda_0,\theta_0)$ for $t_\pi=\pi/(2g_{\rm p})=0.714879~\mathrm{ms}$.
The pair channel transfers population from $\lvert1;gg\rangle$ toward $\lvert0;ee\rangle$. At the
end of this interval, the total dressed double-excitation population is $P_{ee}=0.99580$.

During \emph{freeze}, $\theta=\theta_0$ remains fixed while
$\Lambda(u)=\Lambda_0+(\Lambda_{\rm off}-\Lambda_0)\mathcal R(u)$.
The local dressed states remain unchanged while the pair doublet moves
away from resonance. At the off-resonant endpoint, we define
$g_{\rm off}\equiv g_{\rm pair}(\Lambda_{\rm off},\theta_0)$ and denote
the detuning $\delta_{{\rm p},1}$ by $\delta_{{\rm p},1,{\rm off}}$ evaluated at $\Lambda=\Lambda_{\rm off}$,
$\theta=\theta_0$, and $\omega_m=\omega_{\rm res}^{(2)}$.
Within the effective pair description, population initially in one
pair state can be transferred to the other with probability
\begin{equation}
P_{\rm rev}(t)
=
\frac{4g_{\rm off}^2}
{\delta_{{\rm p},1,{\rm off}}^2+4g_{\rm off}^2}
\sin^2\!\left[
\frac{t}{2}
\sqrt{\delta_{{\rm p},1,{\rm off}}^2+4g_{\rm off}^2}
\right].
\label{eq:S_freezeBound}
\end{equation}
The corresponding envelope is small when
$|\delta_{{\rm p},1,{\rm off}}|\gg|g_{\rm off}|$.
For $\Lambda_{\rm off}=\Lambda_0+20g_{\rm p}$,
$|\delta_{{\rm p},1,{\rm off}}|/|g_{\rm off}|=40.25$, giving an
envelope of $2.46\times10^{-3}$.
We use $\tau_{\rm freeze}=0.01t_\pi$; the total dressed
$P_{ee}$ after this ramp is $0.99580$.

During \emph{undress}, $\Lambda=\Lambda_{\rm off}$ is fixed and
$\theta(u)=\theta_0[1-\mathcal R(u)]$.
Rotation of the instantaneous dressed basis produces a local
nonadiabatic coupling of order $|\dot\theta|$, so local adiabatic
following requires $|\dot\theta|\ll\Lambda_{\rm off}$. At the same
time, the pair transition must remain outside its avoided crossing
throughout the ramp. With $\Lambda=\Lambda_{\rm off}$ and
$\omega_m=\omega_{\rm res}^{(2)}$ fixed, a sufficient condition is
\begin{equation}
\min_{0\le\theta\le\theta_0}
\left|\delta_{{\rm p},1}\right|
\gg
\max_{0\le\theta\le\theta_0}
\left|g_{\rm pair}(\Lambda_{\rm off},\theta)\right|.
\label{eq:S_guard}
\end{equation}
For the chosen $\Lambda_{\rm off}$, the minimum detuning in
Eq.~\eqref{eq:S_guard} is $25.2$ times the maximum pair-coupling coefficient, and
the smoothstep gives
$\max|\dot\theta|/\Lambda_{\rm off}=4.51\times10^{-3}$.
The single-spin spin--phonon transition also remains off resonant, with
$|\omega_m-\Lambda_{\rm off}|=O(\Lambda_{\rm off})$.
We use $\tau_{\rm undress}=0.10t_\pi\simeq71.49~\mu{\rm s}$.
As $\theta\rightarrow0$, the microwave amplitude vanishes and the
instantaneous dressed states reduce to the bare encoded states, so the
populated pair state evolves into $\lvert0;11\rangle$ without
recrossing the pair resonance.

Denoting the end of the sequence by $t_f$, the complete sequence lasts
$t_f=0.82340~\mathrm{ms}$. The full five-stage propagation uses
$N_{\rm ph}^{\rm prot}=8$ mechanical Fock states. The one-phonon
spectroscopy uses $N_{\rm ph}^{\rm spec}(1)=10$, while the
Fock-sector scaling calculation uses
$N_{\rm ph}^{\rm spec}(n)=\max(10,n+5)$; hence the $n=9$ point in
Fig.~\ref{fig:gating}(b) uses $N_{\rm ph}^{\rm spec}(9)=14$.
For this highest Fock sector, increasing $N_{\rm ph}^{\rm spec}(9)$ from
$14$ to $16$ changes the projected crossing center by only
$8.2\times10^{-8}g_{\rm p}$ and the peak $P_{ee}$ by
$3.9\times10^{-11}$. Repeating the full sequence with
$N_{\rm ph}^{\rm prot}=5,6,8,10$ changes both the final $P_{11}$ and
$\mathcal W/E_{\rm cap}$ by less than $10^{-8}$.
Selected stage-end populations are summarized in
Table~\ref{tab:S_mapping}.

\begin{table}[t]
\caption{
Selected populations along the one-phonon five-stage charging
sequence. $P_{gg}$ and $P_{ee}$ are total dressed-spin populations
summed over the retained mechanical Fock sectors, whereas $P_{0;ee}$
is the population of the specific joint state $\lvert0;ee\rangle$.
The vacuum endpoint is obtained with the same control waveform.
}
\label{tab:S_mapping}
\centering
\small
\begin{tabular}{lc}
\toprule
Quantity & Value \\
\midrule
$P_{gg}$ after dress & $0.99704$ \\
$P_{ee}$ at transfer-stage entry & $7.63\times10^{-4}$ \\
$P_{0;ee}$ at the end of transfer & $0.99577$ \\
Total $P_{ee}$ at the end of transfer & $0.99580$ \\
Total $P_{ee}$ after freeze & $0.99580$ \\
Final $P_{11}$, one-phonon input & $0.999541$ \\
Final $P_{11}$, vacuum input & $7.29\times10^{-5}$ \\
\bottomrule
\end{tabular}
\end{table}

\subsection{Static battery and ergotropy}
\label{sec:S_staticwork}

After undress, the microwave amplitude vanishes and the battery is
defined by the two bare encoded spins. Tracing out the mechanical mode
gives $\rho_B^f=\operatorname{Tr}_m\rho(t_f)$, with fully charged-state
population $P_{11}^f=\langle11|\rho_B^f|11\rangle$.
Its static Hamiltonian is
\begin{equation}
H_B
=
\hbar\omega_{B,1}N_1
+\hbar\omega_{B,2}N_2
+4\hbar JN_1N_2,
\qquad
N_j=\lvert1_j\rangle\!\langle1_j\rvert,
\label{eq:S_HB}
\end{equation}
where $\omega_{B,j}$ are the drive-free bare-spin transition
frequencies; the conditional dipolar contribution is represented
explicitly by the $4\hbar JN_1N_2$ term.
With the discharged state $\lvert00\rangle$ chosen as the energy zero,
$\langle00|H_B|00\rangle=0$ and
$E_{\rm cap}=\langle11|H_B|11\rangle$.
For the reference parameters, the energies of
$\{\lvert00\rangle,\lvert01\rangle,\lvert10\rangle,\lvert11\rangle\}$
in units of $\hbar\Lambda_0$ are
$\{0,3685.3578337,3713.9292623,7399.5899531\}$, so
$E_{\rm cap}=7399.5899531\,\hbar\Lambda_0$.

The final stored energy is
$E_B^f=\operatorname{Tr}(\rho_B^fH_B)$.
To construct the passive state, write
$\rho_B^f=\sum_k r_k\lvert r_k\rangle\!\langle r_k\rvert$ with
$r_1\ge r_2\ge\cdots$ and
$H_B=\sum_k\epsilon_k
\lvert\epsilon_k\rangle\!\langle\epsilon_k\rvert$ with
$\epsilon_1\le\epsilon_2\le\cdots$.
The passive state is
$\pi_B^f=\sum_k r_k
\lvert\epsilon_k\rangle\!\langle\epsilon_k\rvert$, with passive
energy $E_{\rm pas}^f=\operatorname{Tr}(\pi_B^fH_B)$.
The final ergotropy,
$\mathcal W^f=E_B^f-E_{\rm pas}^f$, is the maximum work extractable
from $\rho_B^f$ by a cyclic unitary operation
~\cite{Allahverdyan2004}.

For the one-phonon input,
$P_{11}^f=0.999541$,
$E_B^f/E_{\rm cap}=0.999542$,
$\mathcal W^f/E_{\rm cap}=0.999347$, and
$E_{\rm pas}^f/E_{\rm cap}=1.95\times10^{-4}$.
The Hamiltonian in Eq.~\eqref{eq:S_HB} is also used to evaluate the
battery-energy and ergotropy trajectories shown in
Fig.~\ref{fig:battery}(b).

\subsection{Full-cycle classical bound}

The fixed-operation bound of Sec.~\ref{sec:S_classical} also applies to
the normalized final battery energy. Let $U_f$ be the complete unitary
generated by Eq.~\eqref{eq:S_controlH} under the five-stage waveform,
with the two spins initially in $\lvert00\rangle$. For this application,
the generic effect in Eq.~\eqref{eq:S_genericEffect} is
$M=\langle00|U_f^\dagger
(\mathbb I_m\otimes H_B/E_{\rm cap})U_f|00\rangle$.
Writing $\mathbb I_B$ for the identity on the battery Hilbert space,
$0\le H_B/E_{\rm cap}\le\mathbb I_B$ implies
$0\le M\le\mathbb I_m$, and an arbitrary initial mechanical state
$\rho_m$ gives the normalized response
$\operatorname{Tr}_m(M\rho_m)=E_B^f(\rho_m)/E_{\rm cap}$.
For the vacuum input,
$R_{\rm vac}=E_{B,{\rm vac}}^f/E_{\rm cap}=7.31\times10^{-5}$.
At $\bar n=1$, Eq.~\eqref{eq:S_generalClassicalBound} therefore gives
$E_B^f(\rho_{\rm cl})/E_{\rm cap}\le
\mathcal C_{\bar n=1}(R_{\rm vac})=0.64035$ for every classical
mechanical state, whereas the one-phonon input gives
$E_B^f/E_{\rm cap}=0.999542$.
Since $E_{\rm pas}^f\ge0$ implies $\mathcal W^f\le E_B^f$, the same
value is also a conservative classical upper bound on
$\mathcal W^f/E_{\rm cap}$.

\section{Laboratory-frame energy balance and charging efficiency}
\label{sec:S_ledger}

The five-stage dynamics are propagated with Eq.~\eqref{eq:S_controlH}
in the microwave carrier-rotating frames of
Sec.~\ref{sec:S_fullcycle}. To calculate the work performed by the
microwave drives, we use the corresponding projected laboratory-frame
Hamiltonian $H_{\rm lab}^{\rm proj}(t)$, obtained by reversing the
carrier transformations while retaining the encoded-state projection
and approximations of Sec.~\ref{sec:S_microscopic}. Its time dependence
is set by the microwave amplitudes and carrier phases specified in
Sec.~\ref{sec:S_fullcycle}. For the closed-system evolution, the net
control work is
\begin{equation}
W_{\rm ctrl}
=
\int_0^{t_f}dt\,
\operatorname{Tr}\!\left[
\rho_{\rm lab}(t)\partial_t H_{\rm lab}^{\rm proj}(t)
\right]
=
E_{\rm lab}(t_f)-E_{\rm lab}(0),
\label{eq:S_controlWork}
\end{equation}
where $\rho_{\rm lab}(t)$ is the laboratory-frame state and
$E_{\rm lab}(t)=
\operatorname{Tr}[\rho_{\rm lab}(t)H_{\rm lab}^{\rm proj}(t)]$.

The microwave drives vanish at the beginning and end of the
sequence. Omitting the constant energy offset from the static
oscillator displacement, the endpoint Hamiltonian is
\begin{equation}
H_{\rm end}
=
\hbar\omega_m b^\dagger b
+H_B
+\hbar\lambda(b+b^\dagger)(\tau_1^z+\tau_2^z),
\label{eq:S_endpointEnergy}
\end{equation}
where the spin--spin dipolar energy is included in $H_B$.
These terms are invariant under the carrier-frame rotations,
so their endpoint expectation values follow from the states propagated
with Eq.~\eqref{eq:S_controlH}. We define
$E_m(t)=\hbar\omega_m\operatorname{Tr}[\rho(t)b^\dagger b]$ and
$E_{\rm int}(t)=\hbar\lambda\operatorname{Tr}\{\rho(t)
(b+b^\dagger)(\tau_1^z+\tau_2^z)\}$.
At the endpoints, $E_{\rm lab}=E_B+E_m+E_{\rm int}$.
For the one-phonon input, $E_B^i=0$, $E_m^i=\hbar\omega_m$, and
$E_{\rm int}^i=0$ since ${}_m\!\langle1|b+b^\dagger|1\rangle_m=0$.
Writing $E_m^{\rm loss}\equiv E_m^i-E_m^f$, the endpoint balance gives
\begin{equation}
E_B^f
=
W_{\rm ctrl}+E_m^{\rm loss}-E_{\rm int}^f.
\label{eq:S_ledger}
\end{equation}
The term $-E_{\rm int}^f$ corrects for the residual
spin--mechanical interaction energy. The contributions are listed in
Table~\ref{tab:S_energyLedger}.

\begin{table}[t]
\caption{
Laboratory-frame endpoint energy balance for the one-phonon
five-stage sequence. The interaction entry is reported with the sign
with which it appears in Eq.~\eqref{eq:S_ledger}; fractions are
normalized by the final battery energy $E_B^f$.
}
\label{tab:S_energyLedger}
\centering
\small
\begin{tabular}{lcc}
\toprule
Contribution & Energy & Fraction of $E_B^f$ \\
\midrule
Control work $W_{\rm ctrl}$
& $7394.1927\,\hbar\Lambda_0$
& $99.97287\%$ \\
Mechanical energy loss $E_m^{\rm loss}$
& $2.00603\,\hbar\Lambda_0$
& $0.027122\%$ \\
Endpoint interaction correction $-E_{\rm int}^f$
& $+5.62\times10^{-4}\,\hbar\Lambda_0$
& $+7.59\times10^{-6}\%$ \\
\midrule
Final battery energy $E_B^f$
& $7396.199\,\hbar\Lambda_0$
& $100\%$ \\
\bottomrule
\end{tabular}
\end{table}

The one-phonon balance shows that the microwave drives provide
nearly all of the final battery energy. To determine how the initial
phonon changes the energy supplied by the same microwave controls,
we also evaluate the vacuum input under the unchanged five-stage
waveform. For each quantity $X$ in the energy balance, define
$\Delta_{\rm gate}X\equiv X_{\rm 1ph}-X_{\rm vac}$.
Both inputs have $E_{\rm int}^i=0$, so subtracting their endpoint
balances gives
\begin{equation}
\Delta_{\rm gate}E_B^f
=
\Delta_{\rm gate}W_{\rm ctrl}
+\Delta_{\rm gate}E_m^{\rm loss}
-\Delta_{\rm gate}E_{\rm int}^f.
\label{eq:S_differentialLedger}
\end{equation}
For the two inputs,
$W_{{\rm ctrl},{\rm vac}}/E_{\rm cap}=7.31052\times10^{-5}$ and
$W_{{\rm ctrl},{\rm 1ph}}/E_{\rm cap}=0.99927060$.
Table~\ref{tab:S_differentialEnergyLedger} evaluates the differential
balance in Eq.~\eqref{eq:S_differentialLedger}.

\begin{table}[t]
\caption{
Difference between the one-phonon and vacuum endpoint energy balances
under the same five-stage control waveform. Fractions are normalized
by the battery-energy contrast $\Delta_{\rm gate}E_B^f$.
}
\label{tab:S_differentialEnergyLedger}
\centering
\small
\begin{tabular}{lcc}
\toprule
Contribution
& $\Delta_{\rm gate}X/E_{\rm cap}$
& Fraction of $\Delta_{\rm gate}E_B^f$ \\
\midrule
Control work $\Delta_{\rm gate}W_{\rm ctrl}$
& $0.9991975$
& $99.97287\%$ \\
Mechanical energy loss $\Delta_{\rm gate}E_m^{\rm loss}$
& $2.71119\times10^{-4}$
& $0.027126\%$ \\
Endpoint interaction correction $-\Delta_{\rm gate}E_{\rm int}^f$
& $+3.69720\times10^{-8}$
& $+3.70\times10^{-6}\%$ \\
\midrule
Battery-energy contrast $\Delta_{\rm gate}E_B^f$
& $0.999469$
& $100\%$ \\
\bottomrule
\end{tabular}
\end{table}

The differential balance shows that the change in battery energy
between the vacuum and one-phonon inputs is dominated by the change
in microwave control work. The mechanical-energy contribution is
$2.7\times10^{-2}\%$ of the battery-energy contrast, with a much
smaller endpoint interaction correction. Thus the single phonon
gates the pair-charging channel, while the microwave drives
supply nearly all of the energy stored in the battery.

The endpoint balances determine the net microwave control work.
For the charging-efficiency analysis, we count the accumulated
positive microwave work as input, without offsetting it by negative
work. In the carrier-rotating representation of
Eq.~\eqref{eq:S_controlH}, the carrier-averaged microwave power
associated with $H_{\rm lab}^{\rm proj}(t)$ is
\begin{equation}
\overline P_{\rm mw}(t)
=\frac{\hbar\dot\Omega(t)}{2}
\sum_{j=1}^2\langle\tau_j^x\rangle_t
+\hbar\sum_{j=1}^2\omega_{d,j}(t)
\frac{d\langle N_j\rangle_t}{dt}.
\label{eq:S_efficiencyPower}
\end{equation}
The expectation values are taken in the joint spin--mechanical state
$\rho(t)$ evolved under Eq.~\eqref{eq:S_controlH}; $N_j$ and
$\omega_{d,j}(t)$ are defined in Eq.~\eqref{eq:S_HB} and
Sec.~\ref{sec:S_fullcycle}, respectively. We define
\begin{equation}
W_{\pm}(t)=\int_0^t dt'\,
[\pm\overline P_{\rm mw}(t')]_+,
\qquad [x]_+\equiv\max(x,0),
\label{eq:S_efficiencySplit}
\end{equation}
so $W_+(t)$ and $W_-(t)$ give the accumulated positive work and the
magnitude of accumulated negative work, respectively, after summing
the two microwave contributions. Their difference,
$W_{\rm ctrl}(t)=W_+(t)-W_-(t)$, satisfies
$W_{\rm ctrl}(t_f)=W_{\rm ctrl}$.

For the prepared one-phonon input, we include its initial energy
$E_m^i=\hbar\omega_m$, together with $W_+(t_f)$, in the energy
input. In contrast, $E_m^{\rm loss}$ in Eq.~\eqref{eq:S_ledger}
measures the mechanical-energy contribution to the final battery
energy. The charging efficiency is therefore
\begin{equation}
\eta_{\rm ch}
\equiv\frac{\mathcal W^f}{E_m^i+W_+(t_f)}
=\frac{7394.7562}{2.0075+8682.3596}
\simeq85.15\%,
\label{eq:S_chargingEfficiency}
\end{equation}
where the numerical energies are in units of $\hbar\Lambda_0$.

\begin{figure}[htbp]
\centering
\includegraphics[width=0.6\columnwidth]{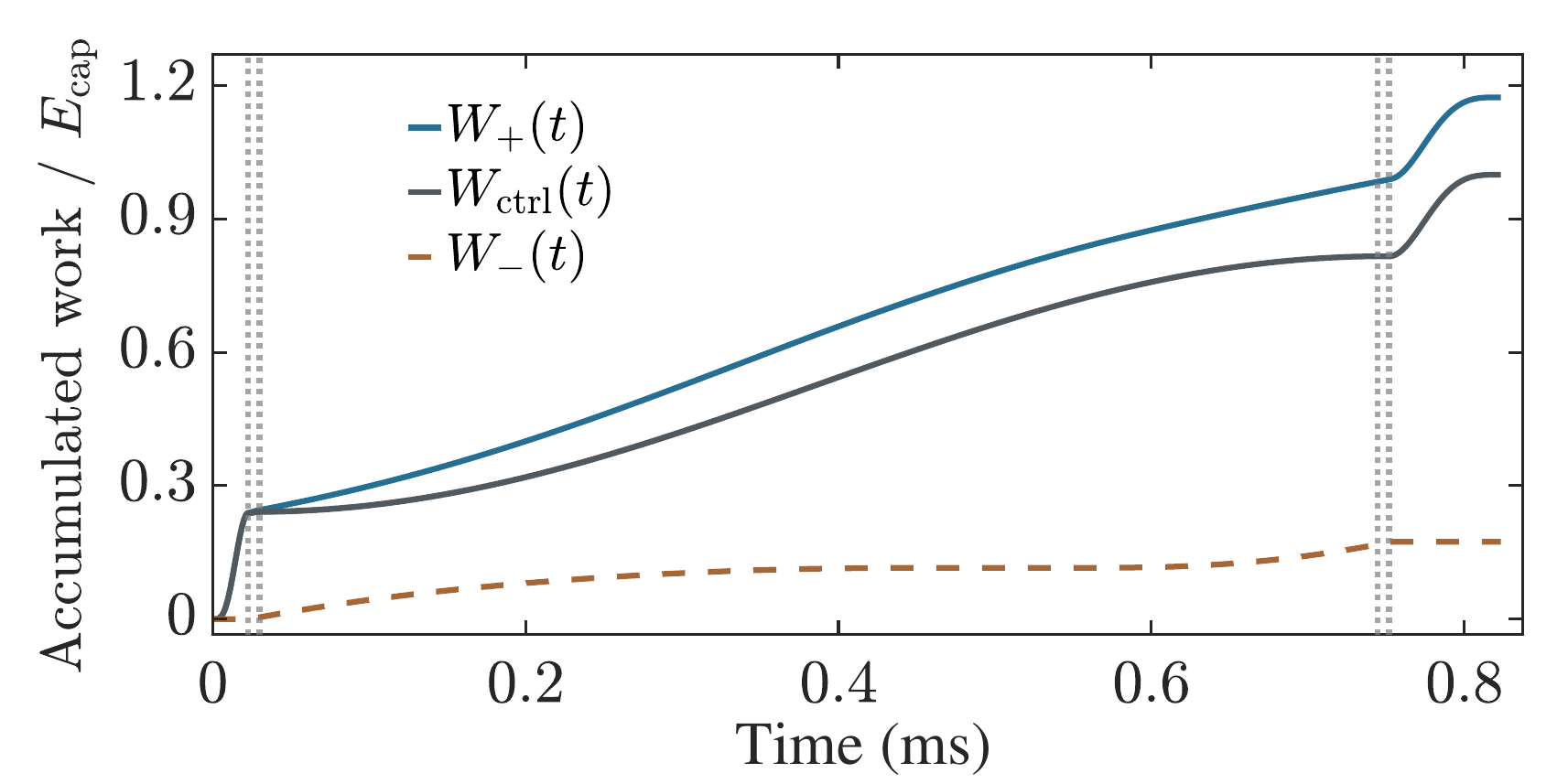}
\caption{
Accumulated microwave work during the five-stage charging sequence
for the one-phonon input $\lvert1;00\rangle$.
The accumulated positive work $W_+(t)$, the accumulated magnitude
of negative work $W_-(t)$, and the net control work $W_{\rm ctrl}(t)=W_+(t)-W_-(t)$ are normalized
by $E_{\rm cap}$. Dotted vertical lines mark the boundaries between stages.
}
\label{fig:S_efficiency}
\end{figure}

Figure~\ref{fig:S_efficiency} shows that approximately $94\%$
of $W_-(t_f)$ accumulates during transfer, where
$(\Lambda,\theta)=(\Lambda_0,\theta_0)$ remains fixed.
The microwave drives remain on, and $\dot\Omega=0$ reduces
Eq.~\eqref{eq:S_efficiencyPower} to
$\overline P_{\rm mw}=\hbar\sum_j\omega_{d,j}
\,d\langle N_j\rangle_t/dt$.
Rapid variations in the bare-spin excitation populations
$\langle N_j\rangle_t$ therefore produce positive and negative
microwave power while the dressed
double-excitation population $P_{ee}$ rises toward its first maximum.

The energy counted as input can be compared directly with the final
ergotropy using Eq.~\eqref{eq:S_ledger} and
$\mathcal W^f=E_B^f-E_{\rm pas}^f$:
\begin{equation}
\begin{aligned}
E_m^i+W_+(t_f)-\mathcal W^f
={}&W_-(t_f)+E_{\rm pas}^f\\
&+E_m^f+E_{\rm int}^f.
\end{aligned}
\label{eq:S_efficiencyPartition}
\end{equation}
The accumulated negative work accounts for $14.83\%$ of the input,
while the signed sum of the remaining endpoint energies contributes
$0.0166\%$. Thus the difference between the input and the final
ergotropy is dominated by the energy exchanged back with the
microwave drives during the coherent evolution.

\section{Operating-window scans}
\label{sec:S_operating}

The operating-window calculations retain the five-stage architecture
and reference stage durations while varying either the mechanical
frequency and transfer hold time or the dressed controls as specified
below. In both cases, the projected propagation returns the final one-phonon ergotropy
$\mathcal W/E_{\rm cap}$, which forms the color scale of
Fig.~\ref{fig:operating}.

The frequency--time scan uses
$\omega_m=\omega_{\rm res}^{(2)}+\Delta\omega_m$ and
$t_{\rm hold}=t_\pi+\Delta t$, with
$-4\le\Delta\omega_m/g_{\rm p}\le4$ and
$-t_\pi\le\Delta t\le t_\pi$. Each selected mechanical frequency is
used throughout the complete five-stage propagation. The dress, tune,
freeze, and undress controls and durations retain their reference
values, while the transfer stage lasts $t_{\rm hold}$. The resulting frequency--time map is shown in
Fig.~\ref{fig:operating}(a).

The control-parameter scan varies the transfer-stage dressed splitting
and mixing angle over $0\le\theta\le\pi$ and
$-4\le(\Lambda-\Lambda_0)/g_{\rm p}\le4$. The complete trajectories
are shifted according to
$\Lambda_{\rm scan}(t)=\Lambda_{\rm ref}(t)+(\Lambda-\Lambda_0)$ and
$\theta_{\rm scan}(t)=(\theta/\theta_0)\theta_{\rm ref}(t)$, where
$\Lambda_{\rm ref}(t)$ and $\theta_{\rm ref}(t)$ are the five-stage
trajectories defined in Sec.~\ref{sec:S_fullcycle}. Thus the
preparation and post-freeze offsets in $\Lambda$ relative to the
transfer-stage value are preserved, while dress and undress still
begin and end at $\theta=0$. The mechanical frequency remains fixed at
$\omega_{\rm res}^{(2)}$ and every stage duration retains its reference
value, including the transfer duration $t_\pi$. The resulting control-parameter map is shown in
Fig.~\ref{fig:operating}(b).

The dashed contours use the effective two-state pair probability
\begin{equation}
P_{ee}^{\rm eff}(t;\delta_{{\rm p},1},g_{\rm pair})
=
\frac{4g_{\rm pair}^2}{\delta_{{\rm p},1}^2+4g_{\rm pair}^2}
\sin^2\!\left[
\frac{t}{2}\sqrt{\delta_{{\rm p},1}^2+4g_{\rm pair}^2}
\right].
\label{eq:S_operatingPeeGeneral}
\end{equation}
For the frequency--time scan, the second-order one-phonon resonance
gives $\delta_{{\rm p},1}=-\Delta\omega_m$ at fixed
$(\Lambda_0,\theta_0)$; Eq.~\eqref{eq:S_operatingPeeGeneral} is then
evaluated with $g_{\rm pair}=g_{\rm p}$ and $t=t_{\rm hold}$.
For the control-parameter scan, the pair-coupling coefficient is
$g_{\rm pair}(\Lambda,\theta)$ and the pair detuning is
$\delta_{{\rm p},1}=2\Lambda-\omega_{\rm res}^{(2)}
+(J^2/\Lambda)(4\cos^2\theta\sin^2\theta+\sin^4\theta)
-(8/3)(\lambda^2/\Lambda)\sin^2\theta$, the same explicit
second-order detuning used in Eq.~\eqref{eq:S_resonance}. By the
definition of $\omega_{\rm res}^{(2)}$, this detuning vanishes at
$(\Lambda_0,\theta_0)$, and the contour is evaluated at $t=t_\pi$.

At fixed $\Lambda$ and $\omega_m=\omega_{\rm res}^{(2)}$, the effective
probability is $\pi$ periodic in $\theta$:
$g_{\rm pair}(\Lambda,\theta+\pi)=-g_{\rm pair}(\Lambda,\theta)$,
whereas the second-order detuning depends only on $\sin^2\theta$ and
$\cos^2\theta$. Hence
$P_{ee}^{\rm eff}(\Lambda,\theta+\pi)
=P_{ee}^{\rm eff}(\Lambda,\theta)$, and
$0\le\theta\le\pi$ spans one full period of the effective prediction.

\section{Validity of the reduced description}
\label{sec:S_validity}

The description used in the Letter involves two successive
reductions with different domains of validity.
First, the electronic ground manifold of each NV center is reduced
to the addressed
$m_s=0,-1$ transition after the microwave and individual-carrier
secular approximations.
Equation~\eqref{eq:centralH} is then diagonalized or propagated directly within this
projected spin--mechanical Hilbert space.
Second, the off-resonant dressed states are eliminated perturbatively
to obtain Eq.~\eqref{eq:pairH}. The projected spectra and time
evolution are obtained from Eq.~\eqref{eq:centralH}, while the $n=1$
pair Hamiltonian gives $\omega_{\rm res}^{(2)}$, $t_\pi$, and the
analytic operating-window contours.

For the microwave rotating-wave approximation, the instantaneous
scale separation
$\Omega_j(t)/\omega_{d,j}(t)\ll1$
is maintained throughout the control waveform, and the
microsecond-scale amplitude and carrier-frequency modulations are slow
compared with the GHz carrier oscillation.
The individual-carrier reduction of the magnetic dipole--dipole
interaction additionally requires
$d_{\rm dd}/\Delta_{\rm enc}\ll1$, where
$\Delta_{\rm enc}\equiv|\omega_{{\rm enc},1}-\omega_{{\rm enc},2}|$
is the separation between the two addressed encoded transitions.
The encoded two-level description requires the drive and interaction
scales to remain small compared with the detuning to the spectator
$m_s=+1$ state.
For the analytic pair reduction, we further require
$|J|/\Lambda\ll1$ and
$|\lambda|/\Lambda\ll1$, while the pair detuning is brought into the
near-degenerate window
$|\delta_{{\rm p},n}|\lesssim|g_n|$.
At the same operating point, the ordinary one-phonon single-spin
transition remains off resonant by
$|\omega_m-\Lambda|=O(\Lambda)$.

At the reference point, we define
$\delta_{\rm ref}\equiv\Lambda_0\cos\theta_0$ and
$\Omega_{\rm ref}\equiv\Lambda_0\sin\theta_0$.
Their values are
$\delta_{\rm ref}/2\pi=0.40415~\mathrm{MHz}$ and
$\Omega_{\rm ref}/2\pi=0.57155~\mathrm{MHz}$, while the two addressed encoded
transitions are separated by
$\Delta_{\rm enc}/2\pi=20.0~\mathrm{MHz}$.
For the transverse dipolar geometry,
$d_{\rm dd}=4J$ and hence
$d_{\rm dd}/2\pi=212~\mathrm{kHz}$, within the range of bare
NV--NV dipolar couplings measured in strongly coupled NV
pairs~\cite{Lee2023}.
For the remaining reference-point scale comparisons,
$\omega_{d,j}$ denotes the carrier frequency corresponding to
$(\Lambda_0,\theta_0)$, and we define
$\omega_{d,\min}\equiv\min_j\omega_{d,j}$.
For the complete-$S=1$ model, the spectator detuning of center
$j$ is
\begin{equation}
\Delta_{+,j}
\equiv
D+\gamma_eB_j-\omega_{d,j},
\qquad
\Delta_{+,\min}
\equiv
\min_j|\Delta_{+,j}|,
\label{eq:S_spectatorDetuning}
\end{equation}
for the electronic model used here.
At the reference point,
$\Delta_{+,\min}/2\pi\simeq0.541~\mathrm{GHz}$.
The resulting dimensionless scale ratios are summarized in
Table~\ref{tab:S_scaleHierarchy}.

\begin{table}[t]
\caption{
Dimensionless scale ratios at the reference point.
For quantities that differ between the two addressed centers, the
smaller denominator is used, so the listed entry is the larger of the
two center-specific ratios.
}
\label{tab:S_scaleHierarchy}
\centering
\small
\begin{tabular}{lc}
\toprule
Ratio & Value \\
\midrule
$\Omega_{\rm ref}/\omega_{d,\min}$
& $2.2\times10^{-4}$ \\

$d_{\rm dd}/\Delta_{\rm enc}$
& $1.06\times10^{-2}$ \\

$\Omega_{\rm ref}/\Delta_{+,\min}$
& $1.1\times10^{-3}$ \\

$J/\Lambda_0$
& $7.57\times10^{-2}$ \\

$\lambda/\Lambda_0$
& $8.57\times10^{-3}$ \\

$g_{\rm p}/\Lambda_0$
& $5.00\times10^{-4}$ \\

$|\omega_{\rm res}^{(2)}-\Lambda_0|/\Lambda_0$
& $1.008$ \\
\bottomrule
\end{tabular}
\end{table}

The largest interaction ratio relevant to the perturbative pair
reduction is $J/\Lambda_0\simeq0.076$. Accordingly, the second-order
pair Hamiltonian is used to identify the virtual mechanism and its
parameter dependence, while all reported projected spectra and
time-dependent numerical results are obtained by direct diagonalization
or propagation of Eq.~\eqref{eq:centralH}.

We next test the local two-level projection by restoring the spectator
electronic state while retaining the carrier-secular reduction of the
dipolar interaction; we call the resulting description the
complete-$S=1$ model.
Let
$\mathbb I_{j,S=1}$ denote the identity on the complete electronic
ground manifold
$\{\lvert m_s=0\rangle_j,
\lvert m_s=-1\rangle_j,
\lvert m_s=+1\rangle_j\}$.
The individual-carrier transformation separates the two eigenspaces of
$(S_j^z)^2$. We denote their spectral projectors by
$\Pi_{j,q}$ with $q=0,1$, so that
$\Pi_{j,0}=\lvert0_j\rangle\!\langle0_j\rvert$ and
$\Pi_{j,1}=\mathbb I_{j,S=1}-\Pi_{j,0}$.
Writing the microscopic dipolar Hamiltonian as
$H_{\rm dd}=\hbar d_{\rm dd}[\bm S_1\!\cdot\!\bm S_2
-3(\bm S_1\!\cdot\!\hat{\bm r}_{12})(\bm S_2\!\cdot\!\hat{\bm r}_{12})]$,
the carrier-secular dipolar interaction in the individual-carrier frame is
\begin{equation}
H_{\rm dd}^{\rm sec}
=
\sum_{q_1,q_2=0}^{1}
(\Pi_{1,q_1}\otimes \Pi_{2,q_2})
H_{\rm dd}
(\Pi_{1,q_1}\otimes \Pi_{2,q_2}).
\label{eq:S_dipolarRWA}
\end{equation}
Equation~\eqref{eq:S_dipolarRWA} removes terms that connect different
carrier sectors of either center while retaining the full
spin-one structure within each sector, including the spectator
$m_s=+1$ state.

After the same microwave rotating-wave reduction, the
carrier-resolved Hamiltonian on the complete ground manifold is
\begin{equation}
\begin{aligned}
\frac{H_{S=1}}{\hbar}
={}&
\omega_m b^\dagger b
+\sum_{j=1}^{2}
\Bigg[
(D-\omega_{d,j})(S_j^z)^2
+\gamma_eB_jS_j^z
+\frac{\gamma_eB_{{\rm mw},j}}{2}S_j^x  \\
&\hspace{30mm}
+\lambda_{S,j}(b+b^\dagger)S_j^z
\Bigg]
+\frac{H_{\rm dd}^{\rm sec}}{\hbar}.
\end{aligned}
\label{eq:S_completeS1}
\end{equation}
The microwave phases have been chosen so that the local drives lie
along the $x$ axis, as in the encoded Hamiltonian.
Restricting the local part of
Eq.~\eqref{eq:S_completeS1} to the
$m_s=0,-1$ block reproduces the addressed transition and its Rabi
coupling, whereas the $m_s=+1$ level remains explicitly present at
the detuning given by Eq.~\eqref{eq:S_spectatorDetuning}.
For the transverse geometry used throughout,
$d_{\rm dd}=4J$, which fixes the dipolar scale consistently between
the projected and complete-$S=1$ descriptions.

Because the spectator state shifts the pair crossing, we locate the
projected and complete-$S=1$ crossing centers independently before
comparing them. For the one-phonon avoided crossing, direct
diagonalization gives
\begin{equation}
\frac{\omega_{\rm res}}{\Lambda_0}
=
2.00748424,
\qquad
\frac{\omega_{\rm res}^{S=1}}{\Lambda_0}
=
2.00772377.
\label{eq:S_S1Resonances}
\end{equation}
The first value is obtained from the projected model, and the second
from the carrier-resolved complete-$S=1$ model.
The spectator manifold therefore shifts the projected-model crossing
by $2.39529\times10^{-4}\Lambda_0$, or $0.4795g_{\rm p}$.
Relative to the second-order one-phonon resonance $\omega_{\rm res}^{(2)}$,
the complete-$S=1$ crossing is displaced by $0.4220g_{\rm p}$. These
shifts are small on the scale of $\Lambda_0$ but are resolved on the
scale of the narrow pair avoided crossing. At its crossing center, the projected model has an avoided-crossing
half-gap of $0.989983g_{\rm p}$ and a mean pair-subspace weight of
$0.996690$.
At its own crossing center, the complete-$S=1$ model has a
half-gap of $0.989867g_{\rm p}$ and a mean pair-subspace weight of
$0.996623$.
Thus the spectator manifold primarily shifts the resonance position
while leaving the pair gap and pair-subspace isolation essentially
unchanged. This static offset can be absorbed by calibrating the transfer-stage
operating point to the observed avoided crossing; the nearly unchanged
half-gap shows that the resonant pair-exchange scale is essentially
preserved after this calibration.
In the complete-$S=1$ spectrum the third-largest pair weight is only
$2.86\times10^{-3}$,
and the nearest other level lies approximately $1.81\times10^3g_{\rm p}$
from the pair center. Repeating the complete-$S=1$ calculation with
mechanical Fock cutoffs of $8$, $10$, and $12$ states changes the crossing
center by less than $5\times10^{-9}g_{\rm p}$ and the half-gap by less than
$3\times10^{-11}g_{\rm p}$.
The two descriptions are therefore compared at their respective
avoided-crossing centers rather than at a common unperturbed value of
$\omega_m$.

The complete-$S=1$ calculation therefore yields a shifted pair
crossing without an additional nearby resonance; the pair doublet
remains isolated.
The five-stage charging dynamics, ergotropy, operating-window maps,
and energy-source balance reported in the Letter are evaluated by
propagating the projected Hamiltonian Eq.~\eqref{eq:centralH}.
Equation~\eqref{eq:pairH} determines the second-order one-phonon resonance
and transfer time and supplies the effective contours used to interpret
those projected-model results.

\section{Robustness to spin decoherence and mechanical thermalization}
\label{sec:S_open}

We next include local Markovian spin noise and mechanical
thermalization in the five-stage charging sequence of
Sec.~\ref{sec:S_fullcycle}. The control waveform and
$H_{\rm enc}(t)$ are unchanged, while the joint spin--mechanical
density operator obeys
\begin{equation}
\begin{aligned}
\dot\rho={}&-\frac{i}{\hbar}[H_{\rm enc}(t),\rho]
+\sum_{j=1}^{2}
\left[
\gamma_1\mathcal D[\tau_j^-]\rho
+\frac{\gamma_z}{2}\mathcal D[\tau_j^z]\rho
\right] \\
&+\kappa(\bar n_B+1)\mathcal D[b]\rho
+\kappa\bar n_B\mathcal D[b^\dagger]\rho ,
\end{aligned}
\label{eq:S_openME}
\end{equation}
where
$\mathcal D[L]\rho=L\rho L^\dagger-\{L^\dagger L,\rho\}/2$,
$\tau_j^-=\lvert0_j\rangle\!\langle1_j\rvert$, and
$\kappa=\omega_m/Q$.
The terms proportional to $\gamma_1$ and $\gamma_z$ describe
encoded-state relaxation and encoded-basis dephasing.
With the coefficient $\gamma_z/2$ in
Eq.~\eqref{eq:S_openME}, an undriven encoded coherence decays as
$e^{-\gamma_z t}$ when $\gamma_1=0$.
The mechanical bath has mean occupation
$\bar n_B=[\exp(\hbar\omega_m/k_BT)-1]^{-1}$.
The one-phonon branch starts from $\lvert1\rangle_m$, whereas the
vacuum branch starts from $\lvert0\rangle_m$; $\bar n_B$ characterizes
the bath coupled to the mechanical mode during the charging sequence.

For fixed noise parameters, Eq.~\eqref{eq:S_openME} generates a
completely positive trace-preserving map $\mathcal E$ from the initial
to the final joint state. For the initial spin state $\lvert00\rangle$,
the final normalized battery energy can be written as
$E_B^f(\rho_m)/E_{\rm cap}
=\operatorname{Tr}_m(M_{\rm open}\rho_m)$, with
$M_{\rm open}=\langle00|\mathcal E^\dagger
(\mathbb I_m\otimes H_B/E_{\rm cap})|00\rangle$.
Because $0\le H_B/E_{\rm cap}\le\mathbb I_B$ and
$\mathcal E^\dagger$ is positive and unital,
$0\le M_{\rm open}\le\mathbb I_m$.
The noisy battery response therefore has the form of
Eq.~\eqref{eq:S_genericEffect}, so the fixed-operation bound of
Sec.~\ref{sec:S_classical} remains valid with
$R_{\rm vac}=E_{B,\rm vac}^f/E_{\rm cap}$ evaluated for the same
noise parameters. Throughout the scans below,
$R_{\rm vac}<e^{-1}$, and the nontrivial branch of
Eq.~\eqref{eq:S_generalClassicalBound} applies at
$\bar n=1$, where $\bar n$ is the mean occupation of the initial
mechanical input. This input-state occupation is distinct from the
bath occupation $\bar n_B$.

Since $\mathcal W^f\le E_B^f$,
$\mathcal C_{\bar n=1}(R_{\rm vac})$ also bounds the normalized
ergotropy attainable from a nonnegative-$P_{\rm GS}$ mechanical input with
$\bar n=1$. We use
$\mathcal G_E=(E_{B,1{\rm ph}}^f-E_{B,\rm vac}^f)/E_{\rm cap}$
to quantify the one-phonon/vacuum charging contrast and
$\Delta_{\rm NC}=\mathcal W_{1{\rm ph}}^f/E_{\rm cap}
-\mathcal C_1(R_{\rm vac})$, with
$\mathcal C_1\equiv\mathcal C_{\bar n=1}$, to quantify the
nonclassical margin. Thus $\Delta_{\rm NC}>0$ rules out every
nonnegative-$P_{\rm GS}$ mechanical input with $\bar n=1$ as a classical
explanation of the one-phonon ergotropy.

The two spin-noise terms in Eq.~\eqref{eq:S_openME} act differently
on the dressed states that participate in pair transfer. On the
transfer plateau, $\theta=\theta_0$ and $\Lambda=\Lambda_0$.
Writing $c=\cos\theta_0$, $s=\sin\theta_0$, and
$\sigma^\pm=(\sigma^x\pm i\sigma^y)/2$, the encoded operators become
$\tau^z=c\sigma^z-s(\sigma^++\sigma^-)$ and
$\tau^-=(s/2)\sigma^z+[(c-1)/2]\sigma^+
+[(c+1)/2]\sigma^-$.
At leading order in the scale separation from $\Lambda_0$, the
zero- and $\pm\Lambda_0$-frequency components of these operators give
\begin{equation}
\begin{aligned}
\mathcal L_z^{\rm dr}
\simeq{}&
\frac{\gamma_zc^2}{2}\mathcal D[\sigma^z]
+\frac{\gamma_zs^2}{2}
\left[
\mathcal D[\sigma^+]+\mathcal D[\sigma^-]
\right],\\
\mathcal L_1^{\rm dr}
\simeq{}&
\frac{\gamma_1s^2}{4}\mathcal D[\sigma^z]
+\frac{\gamma_1(1-c)^2}{4}\mathcal D[\sigma^+]
+\frac{\gamma_1(1+c)^2}{4}\mathcal D[\sigma^-].
\end{aligned}
\label{eq:S_dressedNoise}
\end{equation}
Equation~\eqref{eq:S_dressedNoise} gives the transfer-stage rates
generated by the two encoded-basis dissipators. The full five-stage
propagation below uses Eq.~\eqref{eq:S_openME}.

The first consequence concerns the coherence between the resonant
pair states, $\rho_{\rm p}=\langle1;gg|\rho|0;ee\rangle$.
For a single dressed spin, a term
$a\mathcal D[\sigma^z]$ damps the $g$--$e$ coherence at rate $2a$,
while a population-transfer term
$a\mathcal D[\sigma^\pm]$ contributes $a/2$ to that decay.
Adding the local contributions from the two spins in
Eq.~\eqref{eq:S_dressedNoise} gives
$\Gamma_{{\rm p},z}=\gamma_z(1+c^2)$ for dephasing and
$\Gamma_{{\rm p},1}=\gamma_1(3-c^2)/2$ for relaxation.
At the operating point $c^2=1/3$, these rates become
$\Gamma_{{\rm p},z}=4\gamma_z/3$ and
$\Gamma_{{\rm p},1}=4\gamma_1/3$.
Thus equal values of $\gamma_z$ and $\gamma_1$ produce the same
leading direct decay rate of the resonant pair coherence.

The second consequence concerns dressed-spin excitation in the
vacuum branch. The coefficients of $\mathcal D[\sigma^+]$ in
Eq.~\eqref{eq:S_dressedNoise} give the local
$\lvert g\rangle\rightarrow\lvert e\rangle$ rates
$\Gamma_{\uparrow,z}=\gamma_zs^2/2$ and
$\Gamma_{\uparrow,1}=\gamma_1(1-c)^2/4$.
At $\theta=\theta_0$, they reduce to
$\Gamma_{\uparrow,z}=\gamma_z/3$ and
$\Gamma_{\uparrow,1}\simeq0.0447\gamma_1$.
For equal encoded-basis noise rates, dephasing therefore produces
upward dressed-state transitions about $7.5$ times faster than
relaxation. The vacuum branch enters the transfer plateau close to
$\lvert0;gg\rangle$, so these upward transitions generate spin
excitations without the input phonon that activates the desired pair
transfer. Such excitations can increase
$R_{\rm vac}=E_{B,\rm vac}^f/E_{\rm cap}$ and thereby increase the
classical-input bound $\mathcal C_1(R_{\rm vac})$.
The direct decay of the desired pair coherence and the growth of the
vacuum background therefore probe different effects of spin noise.

Mechanical thermalization introduces a separate timescale.
Applying only the mechanical dissipators in
Eq.~\eqref{eq:S_openME} to $b^\dagger b$ gives
$\left.d\langle b^\dagger b\rangle/dt\right|_{\rm bath}
=-\kappa(\langle b^\dagger b\rangle-\bar n_B)$.
For the vacuum and one-phonon initial states and
$\bar n_B\gg1$, the initial thermal influx is therefore set by
$\kappa\bar n_B$. Over the fixed charging duration $t_f$, the
dimensionless product $\kappa\bar n_Bt_f
=\omega_m\bar n_Bt_f/Q$ measures the accumulated thermal exposure.
Increasing $Q$ reduces this exposure while leaving the prepared
mechanical input unchanged.

\begin{figure}[t]
\centering
\includegraphics[width=0.68\textwidth]{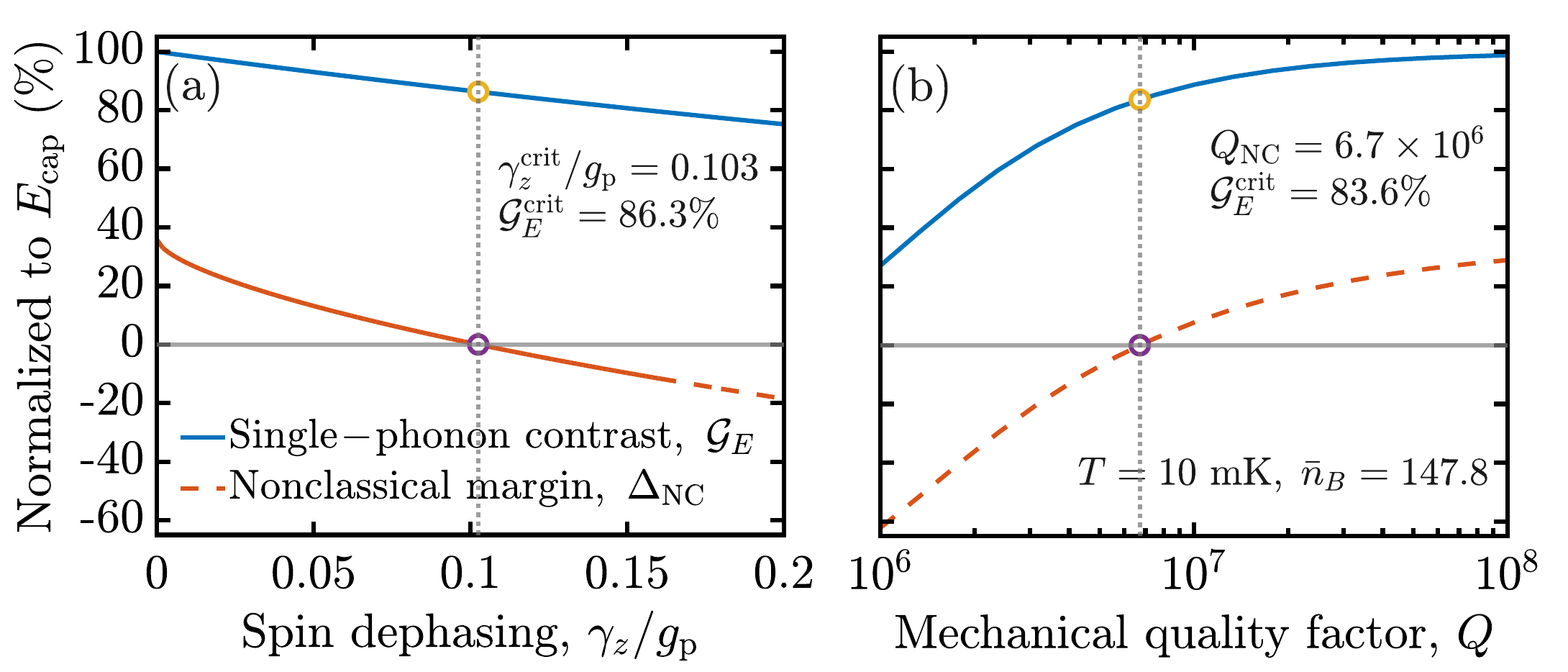}
\caption{
Full five-stage results for separate spin-dephasing and mechanical-bath
scans. Solid curves show the one-phonon/vacuum charging contrast
$\mathcal G_E$, and dashed curves show the nonclassical margin
$\Delta_{\rm NC}$; both are plotted in percent.
(a) Encoded-basis dephasing with $\gamma_1=\kappa=0$.
(b) Mechanical bath at $T=10~\mathrm{mK}$
($\bar n_B=147.8$), with $\gamma_1=\gamma_z=0$.
The horizontal line marks $\Delta_{\rm NC}=0$; the vertical dotted
lines mark the corresponding zero crossings.
}
\label{fig:S_open}
\end{figure}

The full five-stage propagation gives the corresponding noise
thresholds for the complete control waveform. In the dephasing-only
scan of Fig.~\ref{fig:S_open}(a), $\Delta_{\rm NC}$ vanishes at
$\gamma_z^{\rm crit}/g_{\rm p}\simeq0.103$, where
$\mathcal G_E\simeq0.863$. A relaxation-only calculation with
$\gamma_z=\kappa=0$ gives
$\gamma_1^{\rm crit}/g_{\rm p}\simeq0.144$.
The ordering
$\gamma_z^{\rm crit}<\gamma_1^{\rm crit}$ is consistent with the
transfer-stage rates above: at equal $\gamma_z$ and $\gamma_1$ the
direct pair-coherence decay is the same at $\theta_0$, whereas
encoded-basis dephasing produces substantially stronger upward
dressed-state mixing and can therefore raise the vacuum contribution
to $\mathcal C_1(R_{\rm vac})$ more efficiently.

For the mechanical bath,
$\omega_m/2\pi=1.40526~\mathrm{MHz}$ gives
$\bar n_B=147.8$ at $T=10~\mathrm{mK}$.
Figure~\ref{fig:S_open}(b) gives
$Q_{\rm NC}\simeq6.7\times10^6$, for which
$\kappa\bar n_Bt_f\simeq0.160$ and
$\mathcal G_E\simeq0.836$.
At $Q=10^7$,
$\mathcal W_{1{\rm ph}}^f/E_{\rm cap}=0.903$,
$\mathcal G_E=0.887$, and $\Delta_{\rm NC}=0.077$.
At the same quality factor $Q=10^5$, setting
$\bar n_B=0$ and $1$ gives
$\Delta_{\rm NC}=0.309$ and $0.094$.
This comparison isolates the effect of the large thermal occupation:
the higher quality factor required at $10~\mathrm{mK}$ suppresses
repopulation of the mechanical mode from the bath.

The $10$-mK calculations use $N_{\rm ph}=24$.
At $Q=10^6$, the largest change among
$\mathcal W_{1{\rm ph}}^f/E_{\rm cap}$,
$E_{B,\rm vac}^f/E_{\rm cap}$,
$\mathcal G_E$, $\mathcal C_1$, and $\Delta_{\rm NC}$
between $N_{\rm ph}=20$ and $28$ is below
$2.1\times10^{-7}$; between $N_{\rm ph}=24$ and $28$ it is below
$5.6\times10^{-8}$.
Near $Q_{\rm NC}$, changing $N_{\rm ph}$ from $24$ to $28$ changes
these quantities by less than $2.0\times10^{-12}$.
All open-system propagations use relative and absolute tolerances
$10^{-9}$ and $10^{-11}$.
At the dephasing and $10$-mK mechanical thresholds in
Fig.~\ref{fig:S_open}, the charging contrasts remain
$0.863$ and $0.836$.
The strict nonnegative-$P_{\rm GS}$ exclusion therefore closes before the
one-phonon/vacuum charging contrast is strongly reduced.


\begin{thebibliography}{99}

\bibitem{Binder2015}
F.~C.~Binder, S.~Vinjanampathy, K.~Modi, and J.~Goold,
Quantacell: Powerful charging of quantum batteries,
New J. Phys. \textbf{17}, 075015 (2015).

\bibitem{Campaioli2024Review}
F.~Campaioli, S.~Gherardini, J.~Q.~Quach, M.~Polini,
and G.~M.~Andolina,
Colloquium: Quantum batteries,
Rev. Mod. Phys. \textbf{96}, 031001 (2024).

\bibitem{Ferraro2026Perspective}
D.~Ferraro, F.~Cavaliere, M.~G.~Genoni, G.~Benenti,
and M.~Sassetti,
Opportunities and challenges of quantum batteries,
Nat. Rev. Phys. \textbf{8}, 115 (2026).

\bibitem{Allahverdyan2004}
A.~E.~Allahverdyan, R.~Balian, and Th.~M.~Nieuwenhuizen,
Maximal work extraction from finite quantum systems,
Europhys. Lett. \textbf{67}, 565 (2004).

\bibitem{AlickiFannes2013}
R.~Alicki and M.~Fannes,
Entanglement boost for extractable work from ensembles of quantum
batteries,
Phys. Rev. E \textbf{87}, 042123 (2013).

\bibitem{Le2018}
T.~P.~Le, J.~Levinsen, K.~Modi, M.~M.~Parish, and F.~A.~Pollock,
Spin-chain model of a many-body quantum battery,
Phys. Rev. A \textbf{97}, 022106 (2018).

\bibitem{JuliaFarre2020}
S.~Juli\`a-Farr\'e, T.~Salamon, A.~Riera, M.~N.~Bera,
and M.~Lewenstein,
Bounds on the capacity and power of quantum batteries,
Phys. Rev. Research \textbf{2}, 023113 (2020).

\bibitem{Andolina2019}
G.~M.~Andolina, M.~Keck, A.~Mari, M.~Campisi,
V.~Giovannetti, and M.~Polini,
Extractable work, the role of correlations, and asymptotic freedom
in quantum batteries,
Phys. Rev. Lett. \textbf{122}, 047702 (2019).

\bibitem{Shi2022}
H.-L.~Shi, S.~Ding, Q.-K.~Wan, X.-H.~Wang, and W.-L.~Yang,
Entanglement, coherence, and extractable work in quantum batteries,
Phys. Rev. Lett. \textbf{129}, 130602 (2022).

\bibitem{Erdman2024}
P.~A.~Erdman, G.~M.~Andolina, V.~Giovannetti, and F.~No\'e,
Reinforcement learning optimization of the charging of a Dicke
quantum battery,
Phys. Rev. Lett. \textbf{133}, 243602 (2024).

\bibitem{Campaioli2017}
F.~Campaioli, F.~A.~Pollock, F.~C.~Binder, L.~C\'eleri,
J.~Goold, S.~Vinjanampathy, and K.~Modi,
Enhancing the charging power of quantum batteries,
Phys. Rev. Lett. \textbf{118}, 150601 (2017).

\bibitem{Ferraro2018}
D.~Ferraro, M.~Campisi, G.~M.~Andolina, V.~Pellegrini,
and M.~Polini,
High-power collective charging of a solid-state quantum battery,
Phys. Rev. Lett. \textbf{120}, 117702 (2018).

\bibitem{Gyhm2022}
J.-Y.~Gyhm, D.~\v{S}afr\'anek, and D.~Rosa,
Quantum charging advantage cannot be extensive without global
operations,
Phys. Rev. Lett. \textbf{128}, 140501 (2022).

\bibitem{Quach2022}
J.~Q.~Quach, K.~E.~McGhee, L.~Ganzer, D.~M.~Rouse,
B.~W.~Lovett, E.~M.~Gauger, J.~Keeling, G.~Cerullo,
D.~G.~Lidzey, and T.~Virgili,
Superabsorption in an organic microcavity: Toward a quantum battery,
Sci. Adv. \textbf{8}, eabk3160 (2022).

\bibitem{Andolina2018}
G.~M.~Andolina, D.~Farina, A.~Mari, V.~Pellegrini,
V.~Giovannetti, and M.~Polini,
Charger-mediated energy transfer in exactly solvable models for
quantum batteries,
Phys. Rev. B \textbf{98}, 205423 (2018).

\bibitem{Farina2019}
D.~Farina, G.~M.~Andolina, A.~Mari, M.~Polini,
and V.~Giovannetti,
Charger-mediated energy transfer for quantum batteries:
An open-system approach,
Phys. Rev. B \textbf{99}, 035421 (2019).

\bibitem{Rabl2009}
P.~Rabl, P.~Cappellaro, M.~V.~Gurudev Dutt, L.~Jiang,
J.~R.~Maze, and M.~D.~Lukin,
Strong magnetic coupling between an electronic spin qubit and a
mechanical resonator,
Phys. Rev. B \textbf{79}, 041302(R) (2009).

\bibitem{Blais2021}
A.~Blais, A.~L.~Grimsmo, S.~M.~Girvin, and A.~Wallraff,
Circuit quantum electrodynamics,
Rev. Mod. Phys. \textbf{93}, 025005 (2021).

\bibitem{JaynesCummings1963}
E.~T.~Jaynes and F.~W.~Cummings,
Comparison of quantum and semiclassical radiation theories with
application to the beam maser,
Proc. IEEE \textbf{51}, 89 (1963).

\bibitem{Brune1996}
M.~Brune, F.~Schmidt-Kaler, A.~Maali, J.~Dreyer,
E.~Hagley, J.~M.~Raimond, and S.~Haroche,
Quantum Rabi oscillation: A direct test of field quantization in a
cavity,
Phys. Rev. Lett. \textbf{76}, 1800 (1996).

\bibitem{Rinaldi2025}
D.~Rinaldi, R.~Filip, D.~Gerace, and G.~Guarnieri,
Reliable quantum advantage in quantum battery charging,
Phys. Rev. A \textbf{112}, 012205 (2025).

\bibitem{TavisCummings1968}
M.~Tavis and F.~W.~Cummings,
Exact solution for an $N$-molecule---radiation-field Hamiltonian,
Phys. Rev. \textbf{170}, 379 (1968).

\bibitem{Majer2007}
J.~Majer, J.~M.~Chow, J.~M.~Gambetta, J.~Koch,
B.~R.~Johnson, J.~A.~Schreier, L.~Frunzio, D.~I.~Schuster,
A.~A.~Houck, A.~Wallraff, A.~Blais, M.~H.~Devoret,
S.~M.~Girvin, and R.~J.~Schoelkopf,
Coupling superconducting qubits via a cavity bus,
Nature \textbf{449}, 443 (2007).

\bibitem{Bennett2013}
S.~D.~Bennett, N.~Y.~Yao, J.~Otterbach, P.~Zoller,
P.~Rabl, and M.~D.~Lukin,
Phonon-induced spin-spin interactions in diamond nanostructures:
Application to spin squeezing,
Phys. Rev. Lett. \textbf{110}, 156402 (2013).

\bibitem{Li2020}
P.-B.~Li, Y.~Zhou, W.-B.~Gao, and F.~Nori,
Enhancing spin--phonon and spin--spin interactions using linear
resources in a hybrid quantum system,
Phys. Rev. Lett. \textbf{125}, 153602 (2020).

\bibitem{Garziano2016}
L.~Garziano, V.~Macr\`i, R.~Stassi, O.~Di~Stefano,
F.~Nori, and S.~Savasta,
One photon can simultaneously excite two or more atoms,
Phys. Rev. Lett. \textbf{117}, 043601 (2016).

\bibitem{Tomonaga2025}
A.~Tomonaga, R.~Stassi, H.~Mukai, F.~Nori,
F.~Yoshihara, and J.~S.~Tsai,
Spectral properties of two superconducting artificial atoms coupled
to a resonator in the ultrastrong coupling regime,
Nat. Commun. \textbf{16}, 5294 (2025).

\bibitem{Glauber1963}
R.~J.~Glauber,
Coherent and incoherent states of the radiation field,
Phys. Rev. \textbf{131}, 2766 (1963).

\bibitem{Sudarshan1963}
E.~C.~G.~Sudarshan,
Equivalence of semiclassical and quantum mechanical descriptions of
statistical light beams,
Phys. Rev. Lett. \textbf{10}, 277 (1963).

\bibitem{Song2025SelfDischarge}
W.-L.~Song, J.-L.~Wang, B.~Zhou, W.-L.~Yang, and J.-H.~An,
Self-discharging mitigated quantum battery,
Phys. Rev. Lett. \textbf{135}, 020405 (2025).

\bibitem{Doherty2013}
M.~W.~Doherty, N.~B.~Manson, P.~Delaney, F.~Jelezko,
J.~Wrachtrup, and L.~C.~L.~Hollenberg,
The nitrogen-vacancy colour centre in diamond,
Phys. Rep. \textbf{528}, 1 (2013).

\bibitem{PRL136200802}
H.~Gao, N.~T.~Leitao, S.~Dandavate,
L.~B.~Hughes Wyatt, P.~Put, M.~Mammen,
L.~S.~Martin, H.~Park,
A.~C.~Bleszynski Jayich, and M.~D.~Lukin,
Dressed-State Hamiltonian Engineering in
a Strongly Interacting Solid-State Spin Ensemble,
Phys. Rev. Lett. \textbf{136}, 200802 (2026).

\bibitem{Neumann2010}
P.~Neumann, R.~Kolesov, B.~Naydenov, J.~Beck, F.~Rempp,
M.~Steiner, V.~Jacques, G.~Balasubramanian, M.~L.~Markham,
D.~J.~Twitchen, S.~Pezzagna, J.~Meijer, J.~Twamley,
F.~Jelezko, and J.~Wrachtrup,
Quantum register based on coupled electron spins in a room-temperature
solid,
Nat. Phys. \textbf{6}, 249 (2010).

\bibitem{Yamamoto2013}
T.~Yamamoto, C.~M\"uller, L.~P.~McGuinness, T.~Teraji,
B.~Naydenov, S.~Onoda, T.~Ohshima, J.~Wrachtrup, F.~Jelezko,
and J.~Isoya,
Strongly coupled diamond spin qubits by molecular nitrogen implantation,
Phys. Rev. B \textbf{88}, 201201(R) (2013).

\bibitem{SupplementalMaterial}
See Supplemental Material at
[URL will be inserted by publisher]
for the microscopic NV reduction,
effective pair interaction and unequal-cell
generalization, corrected resonance and
finite-coupling effects, classical-input bounds,
complete charging protocol and convergence tests,
laboratory-frame energy balance and charging
efficiency, operating-window construction,
and validity and robustness analyses including
the complete-$S=1$ model, spin decoherence,
and mechanical thermalization.

\bibitem{Bravyi2011}
S.~Bravyi, D.~P.~DiVincenzo, and D.~Loss,
Schrieffer--Wolff transformation for quantum many-body systems,
Ann. Phys. (N.Y.) \textbf{326}, 2793 (2011).

\bibitem{Lee2023}
J. Lee, M. Tatsuta, A. Xu, E. Bauch, M. J. H. Ku, and R. L. Walsworth,
Dressed-state control of effective dipolar interaction between
strongly-coupled solid-state spins,
npj Quantum Inf. \textbf{9}, 77 (2023).

\bibitem{Golter2013}
D.~A.~Golter, K.~N.~Dinyari, and H.~Wang,
Nuclear-spin-dependent coherent population trapping of single
nitrogen-vacancy centers in diamond,
Phys. Rev. A \textbf{87}, 035801 (2013).

\bibitem{Shannon1948}
C.~E.~Shannon,
A Mathematical Theory of Communication,
Bell Syst. Tech. J. \textbf{27}, 379 (1948);
\textbf{27}, 623 (1948).

\bibitem{PhoenixKnight1991}
S.~J.~D.~Phoenix and P.~L.~Knight,
Establishment of an entangled atom-field state in the
Jaynes-Cummings model,
Phys. Rev. A \textbf{44}, 6023 (1991).

\bibitem{Santos2019Stable}
A.~C.~Santos, B.~\c{C}akmak, S.~Campbell, and N.~T.~Zinner,
Stable adiabatic quantum batteries,
Phys. Rev. E \textbf{100}, 032107 (2019).

\bibitem{Niu2024}
Z.~Niu, Y.~Wu, Y.~Wang, X.~Rong, and J.~Du,
Experimental investigation of coherent ergotropy in a single spin
system,
Phys. Rev. Lett. \textbf{133}, 180401 (2024).

\bibitem{Francica2020}
G.~Francica, F.~C.~Binder, G.~Guarnieri, M.~T.~Mitchison,
J.~Goold, and F.~Plastina,
Quantum coherence and ergotropy,
Phys. Rev. Lett. \textbf{125}, 180603 (2020).

\bibitem{dataset_Miao}
Y. Z.~Miao and X. L.~Zhao,
Single-Phonon-Gated Pair Charging of Quantum Batteries data, will be on Zenodo  (2026).

\end{thebibliography}
\end{document}